\documentclass{rQUF2e}
\usepackage{dsfont}
\usepackage{amsmath}
\usepackage[dvipsnames]{xcolor}
\usepackage{algorithmicx}
\usepackage{algorithm} %
\usepackage{algpseudocode}
\usepackage{mycommands}
\usepackage{mathtools} %
\mathtoolsset{showonlyrefs} %

\usepackage[bookmarksopen,pdfstartview=FitH]{hyperref}
\hypersetup{colorlinks,%
	citecolor=blue,%
	filecolor=blue,%
	linkcolor=blue,%
	urlcolor=blue}%
\ifpdf
\hypersetup{
	pdftitle={Algorithmic trading in a microstructural limit order book model},
	pdfauthor={F. Abergel, C. Hur\'e and H. Pham},
	pdfkeywords={MDPs, value iteration, local regression, quantization, limit order book, pure-jump controlled process, market-making, high-dimension}
}
\fi

\usepackage{caption}
\usepackage{subcaption}
\usepackage{tikz}
\usetikzlibrary{fit,matrix,chains,positioning,decorations.pathreplacing,arrows,patterns,shapes.misc}

\theoremstyle{plain}
\newtheorem{theorem}{Theorem}[section]
\newtheorem{corollary}{Corollary}[section]
\newtheorem{lemma}{Lemma}[section]
\newtheorem{proposition}{Proposition}[section]
\newtheorem{definition}{Definition}[section]
\newtheorem{example}{Example}[section]
\newtheorem{remark}{Remark}[section]

\numberwithin{equation}{section}

\newcommand{\red}[1]{#1}

\usepackage{scalerel,stackengine} 

\stackMath
\newcommand\rwh[1]{%
	\savestack{\tmpbox}{\stretchto{%
			\scaleto{%
				\scalerel*[\widthof{\ensuremath{#1}}]{\kern-.6pt\bigwedge\kern-.6pt}%
				{\rule[-\textheight/2]{1ex}{\textheight}}%
			}{\textheight}%
		}{0.5ex}}%
	\stackon[1pt]{#1}{\tmpbox}%
}
\begin{document}

\title{Algorithmic trading in a microstructural limit order book model}  
\author{
	Prof. Fr\'ed\'eric {\sc Abergel}$\dag$
	\qquad
	C\^ome {\sc Hur\'e}$^{\ast}$$\ddag$\thanks{$^{\ast}$corresponding author}
	\qquad
	Prof. Huy\^en {\sc Pham}$\ddag \ddagger$ \\
	\affil{ $\dag$ MICS Laboratory - CentraleSupelec, \href{mailto:frederic.abergel@ecp.fr}{frederic.abergel@ecp.fr} \\
	$\ddag$LPSM - Paris 7 Diderot University, \href{mailto:hure@lpsm.paris}{hure@lpsm.paris} \\
	$\ddag \ddagger$ LPSM - Paris 7 Diderot University and CREST - ENSAE, \href{mailto:pham@lpsm.paris}{pham@lpsm.paris}}
}

\title{Algorithmic trading in a microstructural limit order book model}

\maketitle

\begin{abstract}
	We propose a microstructural modeling framework for studying optimal market-making policies in a FIFO (first in first out) limit order book (order book). In this context, the limit orders, market orders, and cancel orders arrivals in the order book are modeled as point processes with intensities that only depend on the state of the order book. These are high-dimensional models which are realistic from a micro-structure point of view and have been recently developed in the literature. In this context, we consider a market maker who stands ready to buy and sell stock on a regular and continuous basis at a publicly quoted price, and identifies the strategies that maximize their P\&L penalized by their inventory. \red{An extension of the methodology is proposed to solve market-making problems where the orders arrivals are modeled using Hawkes processes with exponential kernel.}
	
	We apply the theory of Markov Decision Processes and dynamic programming method to characterize analytically the solutions to our optimal market-making problem. The second part of the paper deals with the numerical aspect of the high-dimensional trading problem. We use a control randomization method combined with quantization method to compute the optimal strategies. 
	Several computational tests are performed on simulated data to illustrate the efficiency of the computed optimal strategy. In particular, we simulated an order book with constant/ symmetric/ asymmetrical/ state dependent intensities, and compared the computed optimal strategy with naive strategies. Some codes are available on \url{https://github.com/comeh}.
\end{abstract}


\section{Introduction}
Most of the markets use a limit order book (order book) mechanism to facilitate trade. Any market participant can interact with the order book by sending either market orders or limit orders. In such type of markets, the market makers play a fundamental role by providing liquidity to other market participants, typically to impatient agents who are willing to cross the bid-ask spread. The profit made by a market-making strategy comes from the alternation of buy and sell orders.

From the mathematical modeling point of view, the market-making problem corresponds to the choice of an optimal strategy for the placement of orders in the order book. Such a strategy should maximize the expected utility function of the wealth of the market maker up to a penalization of their inventory. 
In the recent litterature, several works focused on the problem of market-making through stochastic control methods.  The seminal paper by  \citet{Av}
inspired by the work of \citet{HoSt} proposes a framework for trading in an order driven market. They modeled a reference price for the stock as a Wiener process, and the arrival of a buy or sell liquidity-consuming order at a distance $\delta$ from the reference price is described by a point process with an intensity in an exponential form decreasing with $\delta$. They characterized the optimal market-making strategies that maximize an exponential utility function of terminal wealth. Since this paper, other authors have worked on related market-making problems. \citet{GuLeFe} generalized the market-making problem of \citet{Av} by dealing with the inventory risk. \citet{CaJa} also designed algorithms that manage inventory risk. \citet{FoPh} and \citet{FoPh2} considered a model designed to be a good compromise between accuracy and tractability, where the stock price is driven by a Markov Renewal Process, and solved the market-making problem. \citet{GiPh} also considered a model for the mid-price, modeled the spread as a discrete Markov chain that jumps according to a stochastic clock, and studied the performance of the market-making strategy both theoretically and numerically. \citet{CaJa2} employed a hidden Markov model to examine the intra-day changes of dynamics of the order book. Very recently, \citet{CaPeJa} and \citet{Gu} published monographs in which they developped models for algorithmic trading in different contexts. \citet{ElAb}  extended the framework of Avellaneda and Stoikov to the options market-making. A common feature of all these works is that a model for the price or/and the spread is considered, and the order book is then built from these quantities. This approach leads to models that predict well the long-term behavior of the order book. The reason for this choice is that it is generally easier to solve the market-making problem when the controlled process is low-dimensional.
Yet, some recent works have introduced accurate and sophisticated micro-structural order book models. These models reproduce accurately the short-term behavior of the market data. The focus is on conditional probabilities of events, given the state of the order book and the positions of the market maker. \citet{Ab} proposed models of order book where the arrivals of orders in the order book are driven by Poisson processes or Hawkes processes. \citet{Co} also modeled the orders arrivals with Poisson processes. \citet{RoLe} proposed a queue-reactive model for the order book. In this model the arrivals of orders are driven by Cox point processes with intensities that only depend on the state of the order book (they are not time dependent). Other tractable dynamic models of order-driven market are available (see e.g. \citet{Co}, \citet{Ro}, \citet{CaJaRi}).

In this paper we adopt the micro-structural model of order book in \citet{Ab}, and solve the associated trading problem.  The problem is formulated in the general framework of Piecewise Deterministic Markov Decision Process (PDMDP), see \citet{Ba}. Given the model of order book, the PDMDP formulation is natural. Indeed, between two jumps, the order book remains constant, so one can see the modeled order book as a point process where the time becomes a component of the state space. As for the control, the market maker fixes their strategy as a deterministic function of the time right after each jump time. 
We prove that the value function of the market-making problem is equal to the value function of an associated non-finite horizon Markov decision process (MDP). This provides a characterization of the value function in terms of a fixed point dynamic programming equation.  \citet{JaLi} recently followed a similar idea to solve an optimal liquidation problem, while \citet{Baretal18} and \citet{Mou18} also tackled this problem of reward functional maximization in a micro-structure model of order book framework.

The second part of the paper deals with the numerical simulation of the value functions. The computation is challenging because the micro-structural model used to model the order book leads to a high-dimensional pure jump controlled process, so evaluating the value function is computationally intensive. We rely on control randomization and Markovian quantization methods to compute the value functions. Markovian quantization has been proved to be very efficient for solving control problems associated with high-dimensional Markov processes. We first quantize the jump times and then quantize the state space of the order book. 
See \citet{Ph}  for a general description of quantization applied to controlled processes. The projections are time-consuming in the algorithm, but Fast approximate nearest neighbors algorithms (see e.g. \citet{MuLo}) can be implemented to alleviate the procedure.
We borrow the values of intensities of the arrivals of orders for the order book simulations from \citet{RoLe}  in order to test our optimal trading strategies.

The paper is organized as follows. The model setup is introduced in Section \ref{secmodel}:  
we present the micro-structural model for the order book, and show how the market maker interacts with the market. 
In Section \ref{existenceCharacterisation}, we prove the existence and provide a characterization of the value function and optimal trading strategies. 
In Section \ref{secnum_ob}, we  introduce a quantization-based algorithm to numerically solve a general class of discrete-time control problem with finite horizon, and then apply it on our trading problem.  
We then  present some results of numerical tests on simulated order book.  Section \ref{sec::hawkes} presents an extension of our model when order arrivals are driven by Hawkes processes, and finally the appendix collects some results used in the paper.

\section{Model setup} \label{secmodel}

\subsection{Order book representation}
We consider a model of the order book inspired by the one introduced in chapter 6 of \citet{Ab}. \\

Let us fix $K\geq 0$. An order book is supposed to be fully described by $K$ \emph{limits}\footnote{\red{limit is also referred to as \emph{quote} or \emph{price level} in the literature.}} on the bid side and $K$ limits on the ask side.  
Denote by $pa_t$ the \emph{best ask}  at time $t$, which is the cheapest price a participant in the market is willing to sell a stock at time $t$, and by $pb_t$ the \emph{best bid} at time $t$, which is the highest price a participant in the market is willing to buy a stock at time $t$.
We use the pair of vectors  $\big(\underline{a}_t,\underline{b}_t\big) = \big(a_t^1,...,a_t^K, b_t^1, \dots , b_t^K \big) $ where
\begin{itemize}
	\item $a^i_t$ is the number of shares available $i$ ticks away from $pb_t$,
	\item -$b^i_t$ is the number of shares available $i$ ticks away from $pa_t$,
\end{itemize}
to describe the order book. 
The vectors $\underline{a}_t$ and $\underline{b}_t $ describe respectively the ask and the bid sides at time $t$.
The quantities $a^i_t$, $1 \leq i \leq K$, live in the discrete space $q\mathbb{N}$ where $q \in \mathbb{R}^*$ is the minimum order size on each specific market (\emph{lot size}).  The quantities $b^i_t$, $1 \leq i \leq K$, live in the discrete space $-q\mathbb{N}$. 
By convention, the $a^i$ are non-negative, and the $b^i$ are non-positive for $0 \leq i \leq K$. The tick size $\epsilon$ represents the smallest intervall between different price levels. 
We assume in the sequel that the orders arrivals have the same size $q=1$, and set the tick size to $\epsilon=1$ for simplicity.

Constant boundary conditions are imposed outside the moving frame of size $2K$ in order to guarantee that both sides of the LOB are never empty: we assume that all the limits up to the $K$-th ones are equal to $a_\infty$ in the ask side, and equal to $b_\infty$ in the bid side, with $a_\infty, - b_\infty \in \N$.
\red{The order book can receive at any time three different kinds of orders from general market participants: market orders, limit orders and cancel orders. The orders arrivals are modeled by the following point processes:}
\red{	\begin{itemize}
		\item $M^+$ stands for the buy market orders flow, and we denote by ${\lambda^M}^+$ its intensity,
		\item $M^-$ stands for the sell market orders flow, and we denote by ${\lambda^M}^-$ its intensity,
		\item $L^+_i$, for $i\in \{1,...K\}$, stands for the sell orders flow at the $i^\text{th} $ limit on the ask side, and we denote by $\lambda^{L^+}_i$ its intensity,
		\item $L^-_i$, for $i\in \{1,...K\}$, stands for the buy orders flow at the $i^\text{th} $ limit on the bid side, and we denote by $\lambda^{L^-}_i$ its intensity,
		\item $C^+_i$, for $i\in \{1,...K\}$, stands for  the cancel orders flow at the $i^\text{th} $ limit on the ask side, and we denote by $\lambda^{C^+}_i$ its intensity,
		\item  $C^-_i$, for $i\in \{1,...K\}$, stands for  the cancel orders flow at the $i^\text{th} $ limit on the bid side, and we denote by $\lambda^{C^-}_i$ its intensity.
	\end{itemize}}
We assume in the sequel that \vspace{2mm}\\
 \textbf{(Harrivals)} The orders arrivals from general market participants (market orders, limit orders and cancel orders) occur according to Markov jump processes which intensities only depends on the couple $\big( \underline{a}, \underline{b} \big)$. 
Moreover, we assume that the all the intensities are at most linear w.r.t. the couple $\big( \underline{a}, \underline{b} \big)$ and are constant between two events.

\noindent Under \textbf{(Harrivals)}, 
Let $\lambda^L$, $\lambda^C$, $\lambda^M$ be positive real constants such that
\begin{align}
\sum_{i=1}^{K}\lambda^{L^+}_i(\underline{a},\underline{b})  + \sum_{i=1}^{K}\lambda^{L^-}_i(\underline{a},\underline{b}) &\leq \lambda^L\big(\abs{\underline{a}}+\abs{\underline{b}}\big),\\
\quad \sum_{i=1}^{K}\lambda^{C^+}_i(\underline{a},\underline{b})  + \sum_{i=1}^{K}\lambda^{C^-}_i(\underline{a},\underline{b}) &\leq \lambda^C \big(\abs{\underline{a}}+\abs{\underline{b}}\big),\\
{\lambda^M}^+(\underline{a},\underline{b}) +{\lambda^M}^-(\underline{a},\underline{b}) &\leq \lambda^M \big(\abs{\underline{a}}+\abs{\underline{b}}\big),
\end{align}
for all state $(\underline{a},\underline{b})$ of the LOB, where $|a|:=\sum_{k=1}^{K}-a^k$ and $|b|:=\sum_{k=1}^{K}b^k$.
\begin{remark}
	The linear conditions on the intensities are required to prove that the control problem is well-posed.
\end{remark}
\begin{remark}
\red{	We assume the intensity to be constant between jumps in \textbf{(Harrivals)} for simplicity. All the results proposed in Section \ref{existenceCharacterisation} can be extended to the case where the intensities of the jump processes are deterministic between jumps. Such an extension is considered in Section \ref{sec::hawkes}, where the arrivals are modeled using Hawkes processes with exponential kernel.}
\end{remark}
\begin{remark}
\red{Some information can be integrated to the order book model by adding new processes. For example, some exogenous processes that send orders to the best-ask and best-bid limits can be added to model the predictions of the mid-price and its volatility that an agent may have. Doing so is critical to manage the risk-reward tradeoffs.}
\end{remark}

\subsection{Market maker strategies}
\red{We assume that the order book matching is done on price/time priority, which means that each limit of the order book is a queue where the first order in the queue is the first one to be executed}\footnote{such an order book is sometimes referred to in the literature as an order book governed by a \emph{FIFO} (\emph{First In First Out}) rule.}.

We consider a market maker who stands ready to send buy and sell limit orders on a regular and continuous basis at quoted prices.
A usual assumption in stochastic control theory to characterize the value function as solution to a HJB equation is to constrain the control space to be compact. In this spirit, we shall make the following assumption: 
\vspace{2mm}\\
\noindent \textbf{(Hcontrol)}
Assume that at any time, the total number of limit orders placed by the marker maker does not exceed a fixed (possibly large) integer $\bar{M}$.

\subsubsection{Control of the market maker}
\label{AS}
The market maker can choose at any time to keep, cancel or take positions in the order book (as long as she does not hold more than $\bar M$ positions in the order book).
Her positions are fully described by the following $\bar M-$dimensional vectors $\underline{ra}_t$, $\underline{rb}_t$, $\underline{na}_t$, $\underline{nb}_t$ where $\underline{ra}$ (resp.  $\underline{rb}$) records the limits in which the market maker's sell (resp. buy) orders are located; and $\underline{na}$ (resp. $\underline{nb}$) records the ranks in the  queues of each market maker's sell (resp. buy) orders. 
In order to guarantee that the strategy of the market maker is predictable w.r.t. the natural filtration generated by the orders arrivals processes, we shall make the following assumption.
\vspace{2mm}\\
\noindent \textbf{(Harrivals2)}
The intensities do not depend on the control. Moreover, the market maker does not cross the spread.

To simplify the theoritical analysis, we also make the following assumption:
\noindent \textbf{(Harrivals3)}
Assume that the market maker does not change their strategy between two orders arrivals of the order book. In other words, the market maker makes a decision right after one of the order arrivals processes $L^\pm, C^\pm,M^\pm$ jumps, and keep it until the next the jump of an order arrival. 
\red{
\begin{remark}
	Assumption \textbf{(Harrivals3)} is mild if the order book jumps frequently, since the market maker can change their decisions frequently in such a case. It can also easily be relaxed by considering piecewise constant controls between jumps (which seems well-adapted to most of the time-discretized control problems met in the industry) or any other parametric family of functions. The results and proofs can then be extended, relying mainly on the PDMDP\footnote{PDMDP stands for Piecewise Deterministic Markov Decision Process, and refers to the the control processes which have deterministic dynamics between (random) jumps.}.
\end{remark}
}

We provide in Figure \ref{exmarket maker} a graphical representation of the controlled LOB. Notice that the market maker interacts with the order book by placing orders at some limits. The latter have ranks that evolve after each orders arrivals.

\begin{figure}[H]
	\centering
	\resizebox{8cm}{!}{
		\begin{tikzpicture}
		\tikzset{cross/.style={cross out, draw=black, minimum size=2*(#1-\pgflinewidth), inner sep=0pt, outer sep=0pt},
			cross/.default={1pt}}
		\tikzset{fleche/.style={->,>=stealth',shorten >=2pt, very thick}}
		\tikzset{dfleche/.style={->,>=stealth',shorten >=2pt,double, double distance = 2pt, very thick}}    
		\draw[->,>=stealth'](-12,0) -- (6,0) node[above right]  {\Huge Price};
		\draw[->] (-12,-8) -- (-12,10) node[ right, text width =5cm ] {\Huge Volume \& Rank in the queues};
		\foreach \x in {-5,...,2} {
			\draw (2*\x ,0.1cm) -- (2*\x, -0.1cm) node[above] {};
		}
		\foreach \x in {-8,...,8} {
			\draw ( -12.1cm,\x ) -- (-11.9cm,\x ) node[left] {\Large \x~ };
		}
		\draw [] (-0.2,0) rectangle (0.2,8);
		\draw [] (2-0.2,0) rectangle (2+0.2,6);
		\draw [] (4-0.2,0) rectangle (4+0.2,5);
		\draw [] (-2-0.2,0) rectangle (-2+0.2,-7);
		\draw [] (-4-0.2,0) rectangle (-4+0.2,-5);
		\draw [] (-6-0.2,0) rectangle (-6+0.2,-6);

		\draw[fleche, draw = red] (-8,4)  to [bend left](-2,-3.5);
		\node[text width= 6cm,above ] at (-8,4) {\Huge {\color{red} Buy Orders \\ of the market maker}};						
		
		\draw [fill=red] (-.2,1) rectangle (.2,2);
		\draw [fill=red] (2-.2,0) rectangle (2.2,1);
		\draw [fill=red] (-2.2,-3) rectangle (-1.8,-4);
		\draw [fill=red] (-6.2,-1) rectangle (-5.8,-2);
		\draw[fleche, draw=red] (4,-4)  to [bend left] (2,0.5);
		\draw[fleche, draw= red] (4,-4) to [bend left] (0,1.5);
		\node[text width= 6cm,right] at (4,-4) { \Huge {\color{red} Sell Orders \\ of the market maker}};

		\draw[dashed, draw =BrickRed] (3.8,5) rectangle  (4.2,6) ;
		\draw[fill =BrickRed] (3.8,10) rectangle  (4.2,11) node[above] {\Huge {\color{BrickRed}New market maker sell order}};
		\draw[dfleche, draw=BrickRed] (4,9.8) to (4,6);
		\draw (-6,-1.5) node[dashed, cross=20pt, line width =.70mm, BrickRed] {\Huge {\color{BrickRed}}};
		\end{tikzpicture}
	}
	\caption{Example of market maker's placements and decisions she might make. In this example: the ask-side of the order book is described by $\underline{a}=(7,5,4)$; and  the bid-side by $\underline{b}=(-6,-4,-5)$. The market maker positions are described by $\underline{ra}=(0,1,-1)$ and $\underline{rb}= (0,2,-1...)$; and the associated ranks vectors are $\underline{na}=(2,1,-1\dots)$ and $\underline{nb}=(4,2, -1)$. We have $i0=0$. \\
		After each order arrival, she can send new limit orders (see action on the top right), cancel some positions (see dashed cross on the bottom left), or just keep their orders unchanged.}
	\label{exmarket maker}
\end{figure}
Denote by $(T_n)_{n\in \mathbb{N}}$ the sequence of jump times of the order book. We denote by $\mathbb{A}$ the set of the admissible strategies, defined as the predictable processes $\big(\underline{ra}_t,\underline{rb}_t\big)_{t \leq T} $ such that the control is constant between two consecutive arrivals of orders from the participants, and such that the order of the market maker do not cross the spread. These conditions reads:
\begin{itemize}
	\item $\text{for all } n \in \mathbb{N},  \big(\underline{ra}_t,\underline{rb}_t \big) \in \{1,...,K\}^{\bar M} \times \{1,...,K\}^{\bar M} \text{ are constant on } \big( T_n,T_{n+1} \big] $
	\item $ra_* , rb_* \geq i0$
\end{itemize}
where, for every vector $\underline{a}$: $ a_* = \Min_{1\leq i \leq K} \{a_i \; s.t.\;\underline{ a}_i \neq -1 \}$; and: \red{$i0= \argmin_{1\leq i \leq K} \big(\underline{a}_i \;s.t.\; a_i>0 \big)$}. The control is the double vector of the positions of the $\bar M$ market maker's orders in the order book. 

By convention, we set in the sequel $ra_i(t)=-1$ if the $i$th market maker's order is not placed in the order book. 

\subsubsection{Controlled order book}
The order book, controlled by the market maker, is fully described by the following state process $Z$: \[Z_t:= \big( X_t,Y_t,\underline{a}_t,\underline{b}_t, \underline{na}_t, \underline{nb}_t,pa_t,pb_t, \underline{ra}_t,\underline{rb}_t \big),\]
where, at time $t$:
\begin{itemize}
	\item $X_t$ is the cash held by the market maker on a zero interest account. 
	\item $Y_t$ is the inventory of the market maker, i.e. it is the (signed) number of shares held by the market maker.
	\item $pa_t$ is the ask price, i.e. the cheapest price a general market participant is willing to sell stock. 
	\item $pb_t$ is the bid price, i.e. the highest price a general market participant is willing to buy stock. 
	\item $\underline{a}_t=(a_1(t),\dots,a_K(t))$ (resp. $\underline{b}_t=(b_1(t),\dots,b_K(t))$) describes the ask (resp. bid) side: $i \in \{1,\dots, K\}$, $a_i(t)$ is the sum of all the general market participants' sell orders which are $i$ ticks away from the bid (resp. ask) price. 
	\item $\underline{ra}_t$ (resp. $\underline{ra}_t$) describes the market maker's orders in the ask (resp. bid) side: for $i \in \{1,..., \bar M\}$,  $\underline{ra}_t(i)$ is the number of ticks between the $i$-th market maker's sell (resp. bid) order and the bid (resp. ask) price. By convention, we set $\underline{ra}_t(i)=-1$ (resp. $\underline{rb}_t(i)=-1$) if the $i$-th sell (resp. buy) order of the market maker is not placed in the order book. As a result $\underline{ra}_t(i), \underline{rb}_t(i) \in \{ 1, \dots , K  \} \cup \{-1\}$.
	\item $\underline{na}_t$ (resp. $\underline{nb}_t$) describes the ranks of the market maker's orders in the ask (resp. bid) side. For $i \in \{1,..., \bar M\}$, $\underline{na}_t(i) \in \big\{-1, ...,  \abs{a}+\bar M\big\}$ (resp. $\underline{nb}_t(i) \in \big\{-1, ...,  \abs{b}+\bar M\big\}$) is the rank of the $i$-th sell (resp. buy) orders of the market maker in the queue. 
	By convention, we assume that $\underline{na}_t(i)=-1$ (resp. $\underline{nb}_t(i)=-1$) if the $i$-th sell (resp. buy) order of the market maker is not placed in the order book.
\end{itemize}

\section{Presentation of the market-making problem. Theoretical resolution.}
\label{existenceCharacterisation}
\subsection{Definition of the market-making problem and well-posedness of the value function}
We denote by $V$ the value function for the following market-making problem:
\begin{equation}
\label{eq1}
V(t,z) =  \underset{ \alpha \in \mathbb{A} }{\sup} \;\; \mathbb{E}_{t,z}^\alpha \left[  \int_{t}^{T} f\big(\alpha_s,Z_s\big) \diff s  +g\big(Z_T^{ } \big) \right], \quad (t,z)\in [0,T] \times E,
\end{equation}		
where: 
\begin{itemize}
	\item $\mathbb{A}$ is the set of the admissible strategies, defined in Section \ref{AS}.
	\item $f$ and $g$ are respectively the instantaneous and terminal reward functions.
	\item $ \mathbb{E}_{t,z}^\alpha$ stands for the expectation conditioned by $Z_t=z$ and when strategy $\alpha=\left(\alpha_s\right)_{t\leq s < T}$ is followed on $[t,T]$.
\end{itemize}

	\begin{example}
	\red{
	The terminal reward $g$ can be defined as the sum of the market maker's terminal wealth function and an inventory penalization term, i.e. $g: z \mapsto x+L(y) - \eta y^2$ where $L$ is the amount earned from the immediate liquidation of the inventory\footnote{$L$ is defined as follows:
		\begin{equation*}
		L(z)= \left\{
		\begin{array}{l l}
		\sum_{k=1}^{\jmath-1}\big[a_k(pa+k\epsilon) \big] + (y-a_0-...-a_{\jmath-1})(pa+\jmath \epsilon) & \text{ if } y<0 \\
		-\sum_{k=1}^{\jmath-1}\big[b_k(pb-k\epsilon) \big]+ (y+b_0+...+b_{\jmath-1})(pb-\jmath \epsilon) & \text{ if } y>0 \\
		0 & \text{ if } y=0,
		\end{array}
		\right.
		\end{equation*}
		for all $z=\big(x,y,\underline{a},\underline{b},\underline{na},\underline{nb},pa,pb, \underline{ra},\underline{rb} \big)$, and where we define:
		\begin{equation*}
		\jmath:= \left\{
		\begin{array}{ll}
		\min \big\{ j \big\vert \sum_{i=1}^{j} a_i > -y \big\} & \text{ if } y<0 \\
		\min \big\{ j \big\vert \sum_{i=1}^{j} \vert b_i \vert >  y \big\} & \text{ if } y>0.
		\end{array}
		\right.
		\end{equation*}}. We remind that $z$ stands for a state of order book; $\eps$ is the tick size of the LOB; $x$ is the value of the risk-free account of the market maker; $\eta$ is the penalization parameter of the market maker; and where we remind that $y$ stands for the (signed) market maker's inventory.}
	
	\red{ 
	The running reward $f$ can stand for a penalization of inventory term: $f(z):=-\gamma y^2$, with $\gamma>0$.
}
\end{example}

We shall assume the following conditions on the rewards to insure the well-posedness of the market-making problem.

\noindent \textbf{(Hrewards)} The expectation of the integrated running reward is uniformly upper-bounded w.r.t. the strategies in $\mathbb{A}$, i.e. 
$$\ssup{\alpha \in \mathbb{A}} \mathbb{E}_{t,z}^\alpha \bigg[\int_{t}^{T} f^+(Z_s, \alpha_s) \diff s \bigg] < +\infty$$ holds; where for all state $z$ and action $a$, we denote $f^+(z,a):=\max(f(z,a),0)$. Moreover, the terminal reward $g(Z_T)$ is a.s. at most linear with respect to the number of events up to time $T$, denoted by $N_T$ in the sequel, i.e. there exists a constant $c_1 >0$ such as $g(Z_T) \leq c_1 N_T$, a.s.. 
\begin{remark}
	Under Assumption \red{\textbf{(Hcontrol)}}, Assumption \textbf{(Hrewards)} holds when $g$ is defined as the wealth of the market maker plus an inventory penalization. In particular, we have $g(Z_T) \leq N_T\bar{M}$, where $\bar{M}$ is the maximal number of orders that can be sent by the market maker, which holds a.s. since the best profit the market maker can make is when their  buy (resp. sell) limit orders are all executed, and then the price keeps going to the right (resp. left) direction. Hence the second condition of \textbf{(Hrewards)} holds with $c_1=\bar{M}$.
\end{remark}

\vspace{2mm}
\noindent The following Lemma \ref{majoreR} tackles the well-posedness of the control problem.
\begin{lemma}
	\label{majoreR}Under \textbf{(Hrewards)} and \textbf{(Hcontrol)}, the value function is well-defined, i.e.
	\[\ssup{\alpha \in \mathbb{A}} \mathbb{E}_{t,z}^\alpha \left[g(Z_T) +\int_{t}^{T} f\big(\alpha_s,Z_s\big) \diff s  \right] < +\infty, \]
	where, as defined previously, $\mathbb{E}_{t,z}^\alpha[.]$ stands for the expectation conditioned by the event $\{Z_t=z\}$, assuming that strategy $\alpha \in \mathbb{A}$ is followed in $[t,T]$.
\end{lemma}

\begin{proof}	
	Denote by $ (N_t)_t$ the sum of all the arrivals of orders up to time $t$. 
	Under \textbf{(Hrewards)}, we can bound $\mathbb{E}_{t,z}^\alpha \left[\int_{t}^{T} f\big(\alpha_s,Z_s\big) \diff s  +g(Z_T) \right]$, the reward functional at time $t$ associated to a strategy $\alpha \in \mathbb{A}$, as follows:
	\begin{align}
	\mathbb{E}_{t,z}^\alpha \left[\int_{t}^{T} f\left(\alpha_s,Z_s\right) \diff s  +g(Z_T) \right] &\leq \ssup{\alpha \in \A} \mathbb{E}^\alpha_{t,z} \left[ g(Z_T)\right] +\ssup{\alpha \in \A} \mathbb{E}^\alpha \left[\int_{t}^{T} f^+(Z_s, \alpha_s) \diff s \right] \nonumber\\
	&\leq  c_1 \ssup{\alpha \in \A} \mathbb{E}^\alpha_{t,0} \left[N_T \right] +\ssup{\alpha \in \A} \mathbb{E}^\alpha_{t,z} \left[\int_{t}^{T} f^+(Z_s, \alpha_s) \diff s \right] \label{boundg},
	\end{align}
	where once again, for all general process $M$ and all $m \in E$, $\mathbb{E}^\alpha_{t,m}[ M_T]$ stands for the expectation of $M_T$ conditioned by $M_t=m$ and assuming that the market maker follows strategy $\alpha \in \mathbb{A}$ in $[t,T]$.
	Let us show that the first term in the r.h.s. of \eqref{boundg} is bounded. On one hand, we have:
	\begin{equation}
	\label{ineq::majoreNt}
	\mathbb{E}^\alpha_{t,0}\left[N_T\right] \leq \normeinf{\lambda } \int_{0}^{T}\E\left( \abs{a}_t + \abs{b}_t \right) \diff t,
	\end{equation} where  $\normeinf{\lambda }:=\lambda^L+\lambda^C+\lambda^M$ is a bound on the intensity rate of $N_t$.
	On the other hand, there exists a constant $c_2>0$ such that $\diff (\abs{a}+\abs{b})_t  \leq c_2 \diff L_t$ so that: $\mathbb{E}^\alpha_{t,\abs{a}_0+\abs{b}_0}\left[\abs{a}_t+\abs{b}_t \right] \leq \abs{a}_0+\abs{b}_0 + c_3 \int_{0}^{t} \mathbb{E}\left[ \abs{a}_s+\abs{b}_s \right] \diff s$.
	Applying Gronwall's inequality, we then get:
	\begin{equation}
	\label{ineq::majoreab}
	\mathbb{E}^\alpha_{t,\abs{a}_0+\abs{b}_0}\left[\abs{a}_t+\abs{b}_t\right] \leq \left( \abs{a}_0+\abs{b}_0 \right)e^{c_3t}.
	\end{equation}
	Plugging \eqref{ineq::majoreab} into \eqref{ineq::majoreNt} finally leads to:
	\begin{equation}
	\mathbb{E}^\alpha_{t,0}\left[N_T\right] \leq c_4e^{c_3T}
	\end{equation}
	wit $c_3$ and $c_4>0$ that do not depends on $\alpha$, which proves that the first term in the r.h.s. of \eqref{boundg} is bounded. Also, its second term in the r.h.s. of \eqref{boundg} is bounded under \textbf{(Hrewards)}.
	Hence, the reward functional is bounded uniformly in $\alpha$, which proves that the value function of the considered market-making problem is well-defined.
\end{proof}

\subsection{ Markov Decision Process formulation of the market-making problem}
\label{sec1}

In this section, we first reformulate the market-making problem as a Markov Decision Process (MDP), and then characterize the value function as solution to a Bellman equation.

Let us denote $(T_n)$ is the increasing sequence of the arrivals of market/limit/cancel order to the market; and let $Z_n:=\phi^{a(Z_{T_n})}(Z_{T_n})$, 	where $\phi^a(z) \in E$ is the state of the order book at time $t$ such that $T_n<t< T_{n+1}$, given that $Z_{T_n}=z$ and given that the strategy $a$ has been chosen by the market maker at time $T_n$.  

Let us consider the Markov Decision Process \red{$(T_n,Z_n)_{n\in \N}$}, which is characterized by the following information $$\underbrace{[0,T]\times E}_{\text{ state space }} \quad, \underbrace{A_z}_{\text{ market maker control }}, \underbrace{\lambda}_{\text{intensity of the jump}}, \underbrace{Q}_{\text{transitions kernel}}, \underbrace{r}_{\text{reward }} $$ where:
\begin{itemize}
	\item\red{$[0,T]\times E$ is the state space of the time-continuous controlled process $(T_n,Z_n)_{n\in \N}$}; and $E:=\R \times \N \times \N^K \times \N^K \times \N^{\bar M} \times \N^{\bar M} \times \N^{\bar M} \times \N^{\bar M}  \times \R \times \R$ is the state space of $(Z_t)$.  For $z\in E$, $z = \big(x,y,\underline{a}, \underline{b}, \underline{na}, \underline{nb}, \underline{ra}, \underline{rb}, pa, pb\big)$ where: $x$ is the cash held by the market maker, $y$ their inventory; \red{$\underline{a}$ and  $\underline{b}$, introduced in Section 2.2.2. represent the orders  in the ask and bid sides of the order book of all the participants except the market maker's}; $\underline{na}$ (resp. $\underline{nb}$) is the $\bar M$-dimensional vector of the ranks of the market maker's sell (resp. buy) orders in the queues ; $\underline{ra}$ (resp. $\underline{rb}$) is the $\bar M$-dimensional vector of the number of ticks the  $\bar M$ market maker's sell (resp. buy) orders are from the bid (resp. ask) price; $pa$ (resp. $pb$) is the ask-price (resp. bid-price).
	\item $A_z$, for every state $z \in E$, is the set of the admissible actions (i.e.the actions the market maker can take) when the order book is at state $z$:
	$$A_z=\Big\{ \underline{ra},\underline{rb} \in \{1,...,K\}^{\bar M} \times \{1,...,K\}^{\bar M} \Big\vert rb_*, ra_* \geq i0 \Big\},$$ where we define $ c_* = \min_{1\leq i \leq K}  \{c_i \vert c_i \ne -1 \}$ and $c0= \argmin_{1\leq i \leq K} \{ \underline{c}_i>0 \}$ for $\underline{c} \in \mathbb{N}^{\bar M}$. We recall that this condition means that the market maker is not allowed to cross the spread. 
	\item $\lambda$ is the intensity of the controlled process $(Z_t)$, and reads:
	\begin{align*}
	\lambda (z) &:= \lambda^{M^+}(z)+\lambda^{M^-}(z) + \sum_{1 \leq j \leq K} \lambda^{L_j^+}(z) +\sum_{1 \leq j \leq K} \lambda^{L_j^-}(z) +\sum_{1 \leq j \leq K} \lambda^{C_j^+}(z) +\sum_{1 \leq j \leq K} \lambda^{C_j^-}(z).
	\end{align*}
	Observe that $\lambda$ does not depend on the strategy $\alpha$ chosen by the market maker since we assumed that the general participants does not "see" the market maker's orders in the order book. Although we wrote $z$ as argument for he intensity of the order book process, it cannot depend on any controlled component variable of the latter. To simplify, the reader can assume that the intensities only depend on the vectors $\underline{a}$ and $\underline{b}$.
	\item $Q$ is the transition kernel of the MDP, which is defined as follows:
	\begin{equation}
	\label{eq:TransitionKernel}
	Q\big(B \times C \vert t,z, \alpha \big):= \lambda(z) \int_{0}^{T-t} e^{-\lambda(z)s} \1_B(t+s)Q'\big(C \vert \phi^\alpha(z),\alpha\big) \diff s + e^{-\lambda(z) (T-t) } \1_{T \in B, z \in C}, 
	\end{equation}
	for all Borelian sets $B \subset \mathbb{R}_+$ and \red{$C \subset E$}, for all $(t,z) \in [0,T]\times E$, for all $\alpha \in A$, and where $Q'$ is the transition kernel of $(Z_t)$ defined for all state $z$ as:
	\[
	Q'\big(z' \vert z,u\big) = 
	\begin{cases}
	\frac{\lambda^{M^+} (z)}{ \lambda(z)} & \text{ if } z'= e^{M^+}(\phi^u(z)) \\
	\hfill \vdots \\
	\frac{\lambda^{C^+}(z)}{ \lambda(z)} &  \text{ if } z'= e^{C^+_K}(\phi^u(z)), \\
	\end{cases}
	\]
where $\phi^u(z)$ is the new state of the controlled order book when decision $u$ as been taken and when the order book was at state $z$ before the decision; $e^{M^+}(z)$ is the new state of the order book right after it received a buy market order, given that it was at state $z$ before the jump; and $e^{C^\pm_i}(z)$ is the new state of the order book right after it received a cancel order from a general market participant on its $i^{th}$ ask/bid limit, given that it was at state $z$. 
\item $r: [0,T] \times E^C  \to \mathbb{R} $ is the running reward associated to the MDP with infinite horizon defined as follows:
\begin{align}
r(t,z,a) &:= -c\big(z,a \big) e^{-\lambda(z)(T-t)}(T-t)\1_{t>T} +c\big(z,a\big) \Big(\frac{1}{\lambda(z)} - \frac{e^{-\lambda(z)(T-t)}}{\lambda(z)} \Big) + e^{-\lambda(z)(T-t)}g(z) \1_{t\leq T},
\label{eq:reward}
\end{align} 
 \red{and its definition is motivated by Proposition \ref{valueeq} below. }
\end{itemize}
The cumulated reward functional associated to the MDP $(T_n,Z_n)_{n\in \N}$ for an admissible policy $(f_n)_{n=0}^\infty$ is defined as: 
\[ V_{\infty, (f_n)}(t,z)  = \mathbb{E}_{t,z}^{(f_n)} \Bigg[\sum_{n=0}^{\infty } r\big(T_n,Z_n, f_n(T_n,Z_n)\big)\Bigg],\]
and the associated value function is the supremum of the cumulated reward functional over all the admissible controls in $\mathbb{A}$, i.e.
\begin{equation}
V_\infty (t,z)= \ssup{(f_n)_{n=0}^\infty \in \mathbb{A}} V_{\infty, \alpha}(t,z) , \hspace{.5cm} (t,z) \in [0,T]\times E, \label{eq::valuefunctionMDP}
\end{equation}
\red{Notice that we used the same notation for admissible controls of the MDP and those of the continuous-time control problem. }

\begin{remark}
	$Q$ is defined as in \eqref{eq:TransitionKernel} because
	\begin{align}
	\mathbb{P}\big(T_{n+1}-T_n\leq t, Z_{n+1} \in B \vert T_0,Z_0,\ldots, T_n,Z_n\big) &= \lambda(Z_n)\int_{0}^{t}e^{-\lambda(Z_n)s} Q'\big(B \vert Z_{T_n}, \alpha_{T_n}\big) \diff s  \\
	&=\lambda(Z_n)\int_{0}^{t}e^{-\lambda(Z_n)s} Q'\big(B \vert Z_{T_n},f_n(Z_n)\big) \diff s ,	\label{eq:motivationQ}
	\end{align}
	holds for any admissible policy $\alpha=(f_n)_{n=0}^\infty \in \mathbb{A}$, for all Borelian $B \subset E$, and for all $t \in [0,T]$.
\end{remark}
In the sequel, we denote $\big([0,T]\times E\big)^C:= \Big\{  \big(t,z,a\big) \in E \times \{1,\dots, K\}^{2 \bar M} \big\vert t \in [0,T], z\in E, a\in A_z \Big\} $,  and
$E^C:= \Big\{  \big(z,a\big) \in E \times \{1,\dots, K\}^{2 \bar M} \big\vert z\in E, a\in A_z \Big\}.$ 
$Q'$ is the stochastic kernel from $E^C$ to $E$ that describes the distribution of the jump goals, i.e.,  $Q'\big(B \vert z,u\big)$ is the probability that the order book jumps in the set $B$ given that it was at state $z\in E$ right before the jump, and the control action $u \in A_z$ has been chosen right after the jump time. 

\begin{remark}
	The MDP is defined in such a way that the control is feedback and constant between two consecutive arrivals of market/limit/cancel orders in the market, i.e. in the time-continuous setting: we restrict ourselves to the control $\alpha = (\alpha_t)$ which are entirely characterized by the decision functions $f_n: [0,T]\times E \to A$, and such that  \[ \alpha_t =f_n(T_n,Z_n) \text{ for  } t \in \big(T_n,T_{n+1}\big] \]
	By abuse of  notation, we denote in the sequel by $\alpha$ the sequence of controls $(f_n)_{n=0}^\infty$.
\end{remark}

The following \red{Proposition \ref{valueeq} motivates the special choice of the running reward $r$ as defined in \eqref{eq:reward}: }
\begin{proposition}
	\label{valueeq} The value function of the MDP defined by \eqref{eq::valuefunctionMDP} coincides with \eqref{eq1}, i.e. we have for all $(t,z) \in E^C$:
	\begin{equation}
	\label{eq:prop1}
	V_\infty (t,z) = V(t,z). 
	\end{equation}
\end{proposition}
\begin{proof}
	\noindent Let us show that for all $\alpha=(f_n) \in \mathbb{A}$ and all $(t,z) \in E^C$
	\begin{equation}
	\label{eq:prop1_step1}
	V_\alpha(t,z) = V_{\infty}^{ (f_n)}(t,z). 
	\end{equation}
	Let us first denote by $H_n:= (T_0,Z_0,...,T_n,Z_n)$.  Notice then that for all admissible strategy $\alpha$:
	\begin{align}
	V_\alpha (t,z) &= \mathbb{E}_{t,z}^\alpha \Bigg[\sum_{n=0}^{\infty} \1_{T>T_{n+1}} \big(T_{n+1} - T_n \big) c\big(Z_n,\alpha_n\big) \\
	&\hspace{2cm}+ \1_{[T_n \leq T < T_{n+1})} \Big( g(Z_T)-\eta {Y_T}^2 + (T-T_n)c\big(Z_n,\alpha_n\big)\Big)  \Bigg]\\
	&= \sum_{n=0}^{\infty } \mathbb{E}_{t,z}^{(f_n)} \Big[r\big(T_n,Z_n,f_n(T_n,Z_n)\big)\Big], \label{eq:valueMDP_CT}
	\end{align}
	where we conditioned by $H_n$ between the first and the second line. We recognize $V_{\infty}^{ (f_n)}$ in the r.h.s. of \eqref{eq:valueMDP_CT}, so that the proof of \eqref{eq:prop1_step1} is completed.
	
	It remains to take the supremum over all the admissible strategies $\mathbb{A}$ in \eqref{eq:prop1_step1} to get \eqref{eq:prop1}.
\end{proof}

\vspace{3mm}

From Proposition \ref{valueeq}, we deduce that the value function of the market-making problem is the same as the value function $V_\infty$ of the discrete-time MDP with infinite horizon. We now aim at solving the MDP control problem. To proceed, we first define the maximal reward mapping for the infinite horizon MDP: 
\begin{align}
(\mathcal{T}v)(t,z)&:=\ssup{a\in A_z} \bigg\{  r(t,z,a) + \int v(t',z')Q(t',z' \vert t,\phi^a(z),a )\bigg\} \nonumber\\
&=\ssup{a \in A_z} \bigg\{  r(t,z,a) + \lambda(z)   \int_{0}^{T-t} e^{-\lambda(z)s}\int v(t+s,z')Q'\big(dz' \vert \phi^a(z), a \big) ds\bigg\}, \label{eqcar}
\end{align}

where we recall that: 
\begin{itemize}
	\item $\phi^\alpha(z)$ is the new state of the order book when the market maker follows the strategy $\alpha$ and the order book is at state $z$ before the decision is taken.
	\item $\lambda(z)$ is the intensity of the order book process given that the order book is at state $z$.
\end{itemize}
\noindent We shall tighten assumption \textbf{(Hrewards)} in order to guarantee existence and uniqueness of a solution to \eqref{eq1}, as well as characterizing the latter.

\vspace{3mm}
\noindent \textbf{(HrewardsBis):} The running and terminal rewards are at most quadratic w.r.t. the state variable, uniformly w.r.t. the control variable, i.e.
\begin{enumerate}
	\item The running reward $f$ is such that $\abs{c}$ is uniformly bounded by a quadratic in $z$ function, i.e. there exists $c_5>0$ such that:
	\begin{equation}
	\forall (z,a) \in E\times A, \quad \abs{f(z,a)} \leq c_5(1+\abs{z}^2).
	\end{equation}		
	\item The terminal reward $g$ has no more than a quadratic growth, i.e. there exists $c_6>0$ such that:
	\begin{equation}
	\forall z \in E, \quad \abs{g(z)} \leq c_6(1+\abs{z}^2).
	\end{equation}
\end{enumerate}

\begin{remark}
	Assumption \textbf{(HrewardsBis)} holds in the case where $g$ is the terminal wealth of the market maker plus a penalization of their inventory, and where with no running reward, i.e. $f=0$.
\end{remark}
\vspace{3mm}
\noindent The main result of this section is the following theorem that gives existence and uniqueness of a solution to \eqref{eq1}, and moreover characterizes the latter as fixed point of the maximal reward operator defined in \eqref{eqcar}.

\begin{theorem}
	\label{proppf}
	$\mathcal{T}$ admits a unique fixed point $v$ which coincides with the value function of the MDP. Moreover we have: \[v= V_\infty=V. \]
	Denote by $f^*$ the maximizer of the operator $\mathcal{T}$.  Then $\big(f^*,f^*,...\big)$ is an optimal stationary (in the MDP sense) policy. 
\end{theorem}
\vspace{2mm}
\begin{remark}
	Theorem \ref{proppf} states that the optimal strategy is stationary in the MDP formulation of the problem, but of course, it is not stationary for the original time-continuous trading problem with finite horizon \eqref{eq1}, since the time component is not a state variable anymore in the original formulation. Actually, given $n \in \mathbb{N}$ and the state of order book $z$ at that time, the optimal decision to take at time $T_n$ is given by $f^*\big(T_n,z \big)$.
\end{remark}

\noindent We devote the next section to the proof of Theorem \ref{proppf}.

\subsection{Proof of Theorem \ref{proppf}}
\label{secSol}

Remind first that we defined in the previous section $E^C:= \Big\{  \big(z,a\big) \in E \times \{1,\dots, K\}^{2 \bar M} \big\vert z\in E, a\in A_z \Big\} $ and  $\big( [0,T] \times E\big)^C:= \Big\{  \big(t,z,a\big) \in [0,T] \times E \times \{1,\dots, K\}^{2 \bar M} \big\vert t \in [0,T], z\in E, a\in A_z \Big\}$.

\begin{definition}
	\label{defBounding}
	A measurable function $b: E \to \mathbb{R}_+$ is called a \emph{bounding function } for the controlled process $(Z_t)$ if there exists positive constants $c_c$, $c_g$, $c_{Q'}, c_\phi$ such that:
	\begin{enumerate}
		\item $\abs{f(z,a)} \leq c_cb(z)$ for all $(z,a) \in E^C$.
		\item $\abs{g(z)} \leq c_gb(z) $ for all $z$ in $E$.  \label{defi2}
		\item $\int b(z') Q'(dz' \vert z,a) \leq c_{Q'}b(z)$ for all $(z,a) \in E^C$.
		\item $b(\phi_t^\alpha (z)) \leq c_\phi b(z) $ for all $(t,z,\alpha ) \in \big( [0,T] \times E\big)^C$.
	\end{enumerate}
\end{definition}

\begin{proposition}
	\label{prop1} 
	Let $b$ be such that :
	$$\forall z \in E, b(z):=1+ \abs{z}^2.$$ 
	Then, $b$ is a bounding function for the controlled process $(Z_t)$, under Assumption \textbf{(HrewardsBis)}.
\end{proposition}
\begin{proof}{} Let us check that $b$ defined in Proposition \ref{prop1} satisfies the four assertions in Definition \ref{defBounding}.
	\begin{itemize}
		\item Assertion 1 and 2 of Definition \ref{defBounding} holds under \textbf{(HrewardsBis)}.
		\item  First notice that $\underline{ra},\underline{rb}$ are bounded by $\sqrt{\bar M}K$ (where we recall that K is the number of limits in each side of the order book, and $\bar M$ is the biggest number of limit orders that the market maker is allowed to send in the market). Secondly, $pa' \in B(pa,K),pb' \in B(pb,K)$, where $B(x,r)$ is the ball centered in $x$ with radius $r>0$, because of the limit conditions that we imposed in our LOB model. And last, we can see that $\abs{\underline{a}'} \leq \abs{\underline{a}}+ a_\infty K$. These three bounds are linear w.r.t. $z$ so that assertion 3 holds.
		\item $\phi^\alpha(z)=z^\alpha$ only differs from $z$ by its $\underline{na}$, $\underline{nb}$, and $\underline{ra}$, $\underline{rb}$ components. \\
		But $\abs{\underline{na}} \leq \sqrt{\bar M}\big(\abs{\underline{a}}+ \bar M\big)$ and $\abs{\underline{nb}} \leq \sqrt{\bar M}\big(\abs{\underline{b}}+\bar M\big)$ are bounded by a linear function of $(\underline{a}, \underline{b})$, also $\abs{\underline{ra}}$ and $\abs{\underline{rb}}$ are bounded by the universal constant $\sqrt{\bar M}K$, so assertion 4 in Definition \ref{defBounding} holds.
	\end{itemize}
\end{proof}
 Let us define \[ \Lambda  :=(4K+2) \ssup{} \left\{\frac{\lambda^{M^\pm}}{\vert \underline{a}\vert + \vert \underline{b}\vert  },\frac{\lambda^{L^\pm}}{\vert \underline{a}\vert + \vert \underline{b}\vert },\frac{\lambda^{C^\pm}}{\vert \underline{a}(z)\vert + \vert \underline{b}(z)\vert } \right\}, \]
which is well-defined under \textbf{(Harrivals)}.
\begin{proposition}
	\label{boundingfunction}
	If $b$ is a bounding function for $(Z_t)$, then  \[ b(t,z):= b(z) e^{\gamma(z) (T-t)} \text{, with } \gamma(z)=\gamma_0 (4K+2) \Lambda  \big(1+\abs{\underline{a}} + \abs{\underline{b}}\big) \text{ and } \gamma_0 >0 \] is a bounding function for the MDP, i.e. for all $t \in [0,T], z \in E,  a \in A_z$, we have: 
	\begin{align}
	\abs{r(t,z,a)} &\leq c_gb(t,z),\\
	\int b(s,z')Q(ds,dz'\vert t,z, a) &\leq c_\phi c_Q e^{C(T-t)}\frac{1}{1+\gamma_0} b(t,z),
	\end{align}
	with $C=\gamma_0\Lambda K(4K+2) \big( \abs{a}_\infty +\abs{b}_\infty \big)$.
\end{proposition}

\begin{proof}
	Let  $z'= \big(x',y',\underline{a}', \underline{b}', \underline{na}', \underline{nb}', \underline{ra}', \underline{rb}'\big)$ be the state of the order book after an exogenous jump \red{occurs} given that it was in state $z$ before the jump. Since $\abs{\underline{a}'} \leq \abs{\underline{a}}+ a_\infty K$ and $\abs{\underline{b}'} \leq \abs{\underline{b}} + b_\infty$, where $a_\infty$ and $b_\infty$ are defined as the border conditions of the order book, we have: 
	\begin{equation}
	\gamma(z') \leq \gamma(z)+ C,
	\label{lemmaj}
	\end{equation}
	with $C=\gamma_0\Lambda K(4K+2) ({a}_\infty +{b}_\infty)$. Then, we get:
	\begin{align*}
	\int b(s,z') Q(ds,dz' \vert t,\phi^\alpha(z),\alpha) &= \lambda(z)   \int_{0}^{T-t} e^{-\lambda(z)s}\int b(t+s,z')Q'\big(dz' \vert \phi_s^\alpha(z), \alpha \big) \diff s 	\\	
	&= \lambda(z)   \int_{0}^{T-t} e^{-\lambda(z)s}\int b(z')e^{\gamma(z')(T-(t+s))} Q'\big(dz' \vert \phi_s^\alpha(z), \alpha \big) \diff s \\
	& \leq \lambda(z)   \int_{0}^{T-t} e^{-\lambda(z)s}\int b(z')e^{(\gamma(z)+C)(T-(t+s))} Q'\big(dz' \vert \phi_s^\alpha(z), \alpha \big) ds\\
	& \leq \lambda(z)   \int_{0}^{T-t} e^{-\lambda(z)s} e^{(\gamma(z)+C)(T-(t+s))} \int b(z') Q'\big(dz' \vert \phi_s^\alpha(z), \alpha \big) ds\\
	& \leq \lambda(z)   \int_{0}^{T-t} e^{-\lambda(z)s} e^{(\gamma(z)+C)(T-(t+s))} c_Qc_\phi b(z) ds\\
	& \leq \frac{\lambda(z) c_Qc_\phi }{\lambda(z)+\gamma(z)+C}  e^{(\gamma (z)+C)(T-t)} \Big(1-e^{-(T-t)(\lambda(z) +\gamma(z) +C)}\Big)b(z) \\
	& \leq c_Qc_\phi \frac{\lambda(z)}{\lambda(z)+\gamma(z)+C}  e^{C(T-t)} \Big(1-e^{-(T-t)(\lambda(z) +\gamma(z) +C)}\Big)b(t,z), 
	\end{align*}
	where we applied \eqref{lemmaj} at the third line.
	It remains to notice that
	\begin{align*}
	\frac{\lambda(z)}{\lambda(z)+\gamma(z)+C} &= \frac{\lambda(z)}{\lambda(z)\big(1+\gamma_0 \big)+ \gamma_0\big[ \underbrace{\Lambda(\abs{a}+\abs{b})-\lambda(z)}_{\geq 0 } \big]  } 
	\leq \frac{1}{1+\gamma_0},
	\end{align*}
	to complete the proof of the proposition. 
\end{proof}

\vspace{3mm}
Let us denote by $\norme{.}_b$ the \emph{weighted supremum norm} such that for all measurable function $v: E' \to \mathbb{R}$,  
\[\norme{v}_b:= \ssup{(t,z) \in E'} \frac{\abs{v(t,z)}}{b(t,z)},\]
and define the set:
\[\mathbb{B}_b:= \Big\{ v: E' \to \mathbb{R} \vert v \text{ is measurable and } \norme{v}_b < \infty \Big\}. \]
Moreover let us define 
\[
\alpha_b:= \ssup{(t,z,\alpha) \in E' \times \mathcal{R}} \frac{\int b(s,z') Q(ds,dz' \vert t,\phi^\alpha(z),\alpha)}{b(t,z)}. 
\]
From the preceding estimations we can bound $\alpha_b$ as follows:
\[
\alpha_b \leq c_Qc_\phi\frac{1}{1 + \gamma_0} e^{CT},
\]
so that, by taking: $\gamma_0=c_Qc_\phi e^{CT}$, we get: $\alpha_b <1$.
In the sequel, we then assume w.l.o.g. that $\alpha_b<1$.
Recall that the maximal reward mapping for the MDP has been defined as:
\[ \mathcal{T}v: (t,z) \mapsto  \ssup{a \in A_z}  \bigg\{ r(t,z,a) + \lambda (z)   \int_{0}^{T-t} e^{-\lambda (z)s}\int v(t+s,z')Q'\big(dz' \vert \phi^a(z), a \big) \diff s \bigg\}
\]
It is straightforward to see that:
\begin{equation}
\label{contracting}
\norme{\mathcal{T}v -\mathcal{T}w}_b \leq \alpha_b \norme{v-w}_b,
\end{equation}
which implies that $\mathcal{T}$ is contracting, since $\alpha_b<1$.

Let $\mathcal{M}$ be the set of all the continuous function in  $\mathbb{B}_b$.
Since $b$ is continuous, $\left(\mathcal{M}, \norme{.}_b \right)$ is a Banach space.

\label{stable}
$\mathcal{T} $ sends $\mathcal{M}$ to $\mathcal{M}$.  Indeed,
for all continuous function $v$ in $\mathbb{B}_b$,
$(t,z,a ) \mapsto r(t,z,a)   + \lambda (z)   \int_{0}^{T-t} e^{-\lambda (z)s}\int v(t+s,z')Q'\big(dz' \vert \phi^a(z), a \big) ds$ is continuous on $[0,T] \times E^C$. $A_z$ is finite, so we get the continuity of the application:
\[ \mathcal{T}v: (t,z) \mapsto  \ssup{a \in A_z}  \bigg\{ r(t,z,a) + \lambda (z)   \int_{0}^{T-t} e^{-\lambda (z)s}\int v(t+s,z')Q'\big(dz' \vert \phi^a(z), a \big) \diff s \bigg\}. 
\]

\begin{proposition}
	\label{prop:maximizer}
	There exists a maximizer for $\mathcal{T}$, i.e. let $v \in \mathcal{M}$, then there exists a Borelian function $f:[0,T] \times E \to A$ such that for all $(t, z) \in E'$: 
	\[\mathcal{T}v\Big(t,z,f\big(t,z\big)\Big) = \ssup{a \in A } \bigg\{ r(t,z,a)+ \lambda(z)   \int_{0}^{T-t} e^{-\lambda(z)s}\int v(t+s,z')Q'\big(dz' \vert \phi^a(z), a \big) \diff s \bigg\} \]
\end{proposition}
\begin{proof}
	$D^*(t,z)= \Big\{ a \in A \big\vert \mathcal{T}_av(t,z)= \mathcal{T}v(t,z)\Big\}$ is finite, so it is compact. So $(t,z) \mapsto D^*(t,z)$ is a compact-valued mapping. 
	Since the application $(t,z,a) \mapsto \mathcal{T}_a(t,z)-\mathcal{T}(t,z)$ is continuous, we get that $D^*= \Big\{ (t,z,a) \in E'^C \big\vert \mathcal{T}_av(t,z)= \mathcal{T}v(t,z)\Big\} $ is borelian. 
	Applying the measurable selection theorem yields to the existence of the maximizer. (see \citet{Ba} p.352)
\end{proof}

\begin{lemma}
	\label{lem::cuthorizon}
	The following holds:
	\begin{equation}
	\ssup{\alpha \in \mathcal{A}} \mathbb{E}_{t,z}^\alpha\left[\sum_{k=n}^{\infty} \abs{r(T_k, Z_k)}\right]  \leq \frac{\alpha^n_b}{1-\alpha_b}b(t,z), \label{convA}
	\end{equation}
	and in particular, we have: 
	\[
	\lim_{n\to \infty} \ssup{\alpha \in \mathcal{A}} \mathbb{E}_{t,z}^\alpha \left[ \sum_{k=n}^{\infty} \left\vert r\big(t_k,Z_k\big) \right\vert \right] = 0.
	\]
\end{lemma}

\begin{proof}
	By conditioning we get $\mathbb{E}_{t,z}^\alpha\Big[\big\vert r(T_k, Z_k) \big\vert \Big] \leq c_g {\alpha_b}^k b(t,z)$ for $k \in \mathbb{N}$, and for all $\alpha \in \mathbb{A}$. It remains to sum this inequality to complete the proof of  Lemma \ref{lem::cuthorizon}.
\end{proof}

\vspace{3mm}
We can now prove Theorem \ref{proppf}. 

\begin{proof}We divided the proof of Theorem \ref{proppf} into four steps.\\
	\noindent \emph{Step 1:}
	Inequality \eqref{contracting} and Proposition \ref{stable} imply that $\mathcal{T}$ is a stable and contracting operator defined on the Banach space $\mathcal{M}$. 
	Banach's fixed point theorem states that $\mathcal{T}$ admits a fixed point, i.e. there exists a function $v \in \mathbb{M}$ such that $v=\mathcal{T} v$, and moreover we have $v= \lim_{n \to \infty} \mathcal{T}^n0$. Notice that $\mathcal{T}^N0$ coincides with $v_0$ defined recursively by the following Bellman equation: 
	\begin{equation}
	\label{eq:bellman1}
	\left\{ 
	\begin{array}{ccl}
	v_N&=&0\\
	v_n&=&\mathcal{T}v_{n+1} \text{ for } n=N-1,...,0.
	\end{array}
	\right.
	\end{equation}
	The solution of the Bellman equation is always larger than the value function of the MDP associated (see e.g. Theorem 2.3.7 p.22 in \citet{Ba}). Then we have:
	$\mathcal{T}^n0 \geq \ssup{(f_k)} \mathbb{E}^{(f_k)}_n\bigg[ \sum_{k=0}^{n-1} r(t_k,X_k) \bigg]  =: J_n$, where $J_n $ is the value function of the MDP with finite horizon $n$ and terminal reward 0, associated to \eqref{eq:bellman1}. Moreover, by Lemma 7.1.4 p.197 in \citet{Ba}, we know that  $\big(J_n\big)_n$ converges as $n \to \infty$ to a limit that we denote by $J$. Passing at the limit in the previous inequality we get: $\lim_{n \to \infty} \mathcal{T}^n0 \geq J$, i.e. 
	\begin{equation}
	\label{eq::1}
	v\geq J.
	\end{equation}
	\nonumber \emph{Step 2:} Let us fix a strategy $\alpha \in \mathbb{A}$, and take $n \in \N$. We denote $J_n(\alpha):= \mathbb{E}^{(\alpha_k)}_0\bigg[ \sum_{k=0}^{n-1} r(t_k,X_k) \bigg] $, the reward functional associated to the control $\alpha$ on the discrete finite time horizon $\{0,\ldots,n\}$. By definition, we have $J_{n}(\alpha)\leq J_n$. We get by letting $n\to \infty$: $\lim_{n \to +\infty} J_{n}(\alpha) =: J_{\infty}(\alpha)\leq J$. Taking the supremum over all the admissible strategies $\alpha$ finally leads to:
	\begin{equation}
	\label{eq::2}
	V_\infty \leq J.
	\end{equation}
	\noindent \emph{Step 3:} Let us denote by $f$ a maximizer of $\mathcal{T}$ associated to $v$, which exists, as stated in Proposition \ref{prop:maximizer}.
	$v$ is the fixed point of $\mathcal{T}$ so that $v= \mathcal{T}^n_f(v)$, for $n\in \N $. Moreover $v \leq \delta$ where $\delta:=\ssup{\alpha\in\Ac}\mathbb{E}\Big[ \sum_{k=0}^{\infty} r^+(Z_k,\alpha_k)\Big]$, so that  $\mathcal{T}^n_f(v) \leq \mathcal{T}^n_f0+ \mathcal{T}^n_o\delta$, where $\mathcal{T}^n_o\delta= \ssup{\alpha} \mathbb{E}^\alpha_{n} \Big[\sum_{k=n}^{\infty} r^+(t_k,Z_k)\Big]$. Lemma \ref{convA} implies that $\mathcal{T}^n_o\delta \to 0$ as $n \to \infty$. Hence, we get: 
	\begin{equation}
	\label{eq::3}
	v \leq J_f.
	\end{equation}
	\noindent \emph{Step 4: Conclusion.} Since
	\begin{equation}
	\label{eq::4}
	J_f \leq V_\infty
	\end{equation}
	holds, we get by combining  \eqref{eq::1}, \eqref{eq::2}, \eqref{eq::3} and \eqref{eq::4}:
	\begin{equation}
	\label{eq::finalresultTheoPrincipal}
	V_\infty \leq J \leq v \leq J_f \leq V_\infty.
	\end{equation}		
	All the inequalities in \eqref{eq::finalresultTheoPrincipal} are then equalities, which completes the proof of Theorem \ref{proppf}. 
\end{proof}

\section{Numerical resolution of the market-making control problem} \label{secnum_ob}

In this section, we first introduce an algorithm to numerically solve a general class of discrete-time control problem with finite horizon, and then apply it on the trading problem \eqref{eq1}.

\subsection{Framework}
Let us consider a general discrete-time stochastic control problem over a finite horizon $N$ $\in$ $\N\setminus\{0\}$.  The dynamics of the controlled state process $Z^\alpha$ $=$ $(Z^\alpha_n)_{n}$ valued in $\R^d$  is given by
\begin{equation}
\label{dynX_ob}
Z_{n+1}^\alpha \; =\;  F(Z_n^\alpha,\alpha_n,\eps_{n+1}), \;\;\; n=0,\ldots,N-1, \; Z_0^\alpha = z \in \R^d, 
\end{equation}
with $(\eps_n)_{n}$  is a sequence of i.i.d. random variables valued in some Borel space $(E,\Bc(E))$, and defined on some probability space $(\Omega,\F,\P)$ equipped with the filtration $\F$ $=$ $(\Fc_n)_n$ generated by the noise $(\eps_n)_n$ ($\Fc_0$ is the trivial $\sigma$-algebra), the control $\alpha$ $=$ $(\alpha_n)_{n}$ is an 
$\F$-adapted process valued in  $A$ $\subset$ $\R^q$, and $F$ is a measurable function from $\R^d\times\R^q\times E$ into $\R^d$. 

Given a running cost function $f$ defined on $\R^d\times\R^q$,  a terminal cost function $g$ defined on $\R^d$, the cost functional associated to a control process  $\alpha$ is
\begin{equation}
J(\alpha) \;= \;\E \left[ \sum_{n=0}^{N-1} f(Z_n^\alpha,\alpha_n) +g(Z_N^\alpha) \right]. 
\end{equation}
The set  $\mathbb{A}$ of admissible controls is the set of control processes $\alpha$ satisfying some integrability conditions  ensuring  that the cost functional $J(\alpha)$ is well-defined and finite. The control problem, also called Markov decision process (MDP), is formulated as  
\begin{equation} \label{defcontrol_ob}
V_0(x_0) \; := \;\sup_{\alpha\in\Ac} J(\alpha),
\end{equation}
and the goal is to find  an optimal control $\alpha^*$ $\in$ $\Ac$, i.e.,  attaining the optimal value: $V_0(z)$ $=$ $J(\alpha^*)$. 
Notice that problem  \eqref{dynX_ob}-\eqref{defcontrol_ob} may also be viewed as the time discretization of a continuous time stochastic control problem, in which case, $F$ is typically the Euler scheme for a controlled diffusion process.

Problem \eqref{defcontrol_ob} is tackled by the dynamic programming approach.
For $n$ $=$ $N,\ldots,0$, the value function $V_n$ at time $n$ is characterized as solution of the following backward (Bellman) equation: 
\begin{equation} \label{DP_ob}
\left\{
\begin{array}{rcl}
V_N(z) & = & g(z)\\
V_n(z) &=& \ssup{a\in A} \left\{ f(z,a) +  \E_{n,z}^a \left[  V_{n+1}(Z_{n+1}) \right] \right\},  \;\;\; z \in \R^d,
\end{array}
\right. 
\end{equation}
Moreover, when the supremum is attained in the DP formula at any time $n$ by $a_n^*(z)$, we get an optimal control in feedback form given by: $\alpha^*$ $=$ 
$(a_n^*(Z_n^*))_n$ where $Z^*$ $=$ $Z^{\alpha^*}$  is the Markov process defined by
\beqs
Z_{n+1}^* &=& F(Z_n^*,a_n^*(Z_n^*),\eps_{n+1}), \;\;\;  n=0,\ldots,N-1, \;\; Z_0^* = z. 
\enqs

There are two usual ways that have been studied in the literature, to solve numerically \eqref{DP_ob}: \red{one way is to use local regression methods, relying e.g. on quantization, k-nearest neighbors or kernel ideas to approximate the conditional expectations by cubature methods}; another way is to rely on MC regress-now or later methods to regress the value functions $V_{n+1}$ at time $n$ for $n=0,\ldots,N-1$ on basis functions or neural networks. See e.g. \citet{lanetal14} for the regress-now and \citet{balata2017} for the regress-later methods for algorithms using basis functions, and e.g. \citet{bacetal18_1} for regression on neural networks based on regress-now or regress-later techniques.

\subsection{Presentation and rate of convergence of the Qknn algorithm}
\label{sec:presQknn}
In this section, we present an algorithm based  on k-nn estimates for local non-parametric regression of the value function, and optimal quantization to quantize the exogenous noise, in order to numerically solve \eqref{DP_ob}. 

\vspace{2mm}
\noindent Let us first introduce some ingredients of the quantization approximation:
\begin{itemize}
	\item We denote by $\hat\eps$ a $K$-quantizer of  the $E$-valued random variable $\eps_{n+1} \sim \eps_1$, 
	that is a discrete random variable on a grid $\Gamma$ $=$ $\{e_1,\ldots,e_K\}$ $\subset$ $E^K$ defined by 
	\begin{equation}
	\hat\eps = {\rm Proj}_\Gamma(\eps_1) \; := \; \sum_{\ell=1}^K e_l 1_{\eps_1 \in C_i(\Gamma)},
	\end{equation}
	where $C_1(\Gamma)$, $\ldots$, $C_K(\Gamma)$ are Voronoi tesselations of $\Gamma$, i.e., Borel partitions of the Euclidian space $(E,|.|)$ satisfying
	\begin{equation}
	C_\ell (\Gamma)  \subset  \left\{ e \in E: |e-e_\ell | \; = \; \min_{j =1,\ldots,K} |e- e_j  | \right\}. 
	\end{equation}
	The discrete law of $\hat\eps$ is then characterized by 
	\begin{equation}
	\hat p_\ell := \P[ \hat\eps = e_\ell ] \; = \; \P[ \eps_1 \in C_\ell (\Gamma) ], \;\;\; \ell=1,\ldots,K.  
	\end{equation}
	The grid points $(e_\ell )$ which minimize the $L^2$-quantization error $\| \eps_1 - \hat\eps\|_{_2}$ lead to the so-called optimal $L$-quantizer, and can be obtained by a stochastic gradient descent method, known as Kohonen algorithm or competitive learning vector quantization (CLVQ) algorithm, which also provides  as a byproduct an estimation of the associated weights $(\hat p_\ell )$. We refer to \citet{Ph} for a description of the algorithm, and mention that for the normal distribution, the optimal grids and the weights of the Voronoi tesselations are precomputed and available on the website 
	\url{http://www.quantize.maths-fi.com}
	\item 
	Recalling the dynamics \eqref{dynX_ob},   the conditional expectation  operator is equal to
	\begin{equation}
	P^{\hat a_n^M(z)} W(x) \; = \; \E\big[ W(Z_{n+1}^{\hat a_n^M}) | Z_n = x \big] = \E \big[ W(F(z,\hat a_n^M(z),\eps_1)) \big], \;\; z \in , 
	\end{equation}
	that we shall approximate analytically by quantization via: 
	\begin{equation} \label{hatquantilater_ob}
	\rwh P^{\hat a_n^M(z)} W(z) := \E \big[ W(F(z,\hat a_n^M(z),\hat\eps)) \big] \; = \; \sum_{\ell=1}^K \hat p_\ell  W(F(z,\hat a_n^M(z),e_\ell )). 
	\end{equation}
\end{itemize}

Let us secondly introduce the notion of  training distribution that will be used to build the estimators of value functions at time $n$, for $n=0,\ldots,N-1$. 
Let us consider a measure $\mu$ on the state space $E$. We refer to it in the sequel as the training measure. Let us take a large integer $M$, and for $n=0,\ldots,N$, introduce $\Gamma_{n}= \left\{ Z^{(1)}_1, \ldots, Z^{(M)}_n \right\}$, where  $\left(Z^{(m)}_n\right)_{m=1}^M$ is a i.i.d. sequence of r.v. following law $\mu$. $\Gamma_{n}$ should be seen as a training sampling to estimate the value function $V_n$ at time $n$.

\vspace{3mm}

The proposed algorithm reads as:
\begin{equation}
\label{algoQknn}
\left\{
\begin{array}{rcl}
\hat{V}_N^Q(z) &=& \quad g(z), \qquad \text{ for } z \in \Gamma_N, \\
\hat{Q}_n(z,a)&=& \quad \sum_{\ell=1}^Kp_\ell \left[ f(z,a) + \hat{V}^Q_{n+1} \left(\mathrm{Proj}_{{n+1}} \left(F\big(z,e_\ell,a\big)\right) \right) \right],  \\
\hat{V}^Q_n(z) &=& \quad \ssup{a\in A} \hat{Q}_n(z,a),\qquad \qquad \text{ for } z \in \Gamma_{n}, \; n=0,\ldots,N-1.
\end{array}
\right.
\end{equation}
where, for $n=0,\ldots,N$, $\mathrm{Proj}_n(z)$ stands for the closest neighbor of $z \in E$ in the grid $\Gamma_n$, i.e. the operator $z \mapsto \mathrm{Proj}_n(z)$ is actually the euclidean projection on the grid $\Gamma_{n}$.

In the sequel, we refer to \eqref{algoQknn} as the Qknn algorithm.

\vspace{3mm}
\noindent We shall make the following assumption on the transition probability of $(Z_n)_{0\leq n \leq N}$, to guarantee the convergence of the Qknn algorithm.

\vspace{2mm}
\noindent \textbf{(Htrans)} Assume that the transition probability ${\mathbb{P}(Z_{n+1} \in A \big\vert Z_n=z,a)}$ conditioned by $Z_n=z$ when control $a$ is followed at time $n$ admits a density $r$ w.r.t. the training measure $\mu$, which is uniformly bounded and lipschitz w.r.t. the state variable $z$, i.e. there exists $\lVert r \rVert_\infty>0$ such that for all $z \in E$ and control $u$ taken at time $n$:
\begin{equation}
\abs{r(y;n,x,a) } \leq \lVert r \rVert_\infty  \quad \text{ and } \quad \abs{r(y;n,x,a) -r(y;n,x',a) } \leq [r]_L\abs{x-x'}
\end{equation}
and $r$ is defined as follows: 
\begin{equation}
\mathbb{P}(Z_{n+1} \in O \big\vert Z_n=z,u) = \int_O r(y; n,x,a) d\mu(y).
\end{equation}
and where we denoted by $[r]_L$ the Lipschitz constant of $r$ w.r.t. $x$.

\vspace{2mm}
Denote by $\mathrm{Supp}(\mu)$ the support of $\mu$. We shall assume smoothness conditions on $\mu$ and $F$ to provide a bound on the projection error. \vspace{2mm}\\
\noindent \textbf{(H$\mu$)} We assume $\mathrm{Supp}(\mu)$ to be bounded, and denote by $\lVert \mu \rVert_\infty$ the smallest real such that $\mathrm{Supp}(\mu) \subset B\left(0,\lVert \mu \rVert_\infty\right)$. Moreover, we assume $x \in E \mapsto \mu \big( B(x,\eta) \big)$ to be Lipschitz, uniformly w.r.t. $\eta$, and we denote by $[\mu]_L$ its Lipschitz constant. 

\vspace{2mm}

\noindent \textbf{(HF)} For $x\in E$ and $a\in A$, assume $F$ to be $\mathbb{L}^1$-Lipschitz w.r.t. the noise component $\eps$, i.e., there exists $[F]_L>0$ such that for all $x\in E$ and $a\in A$, for all r.v. $\eps$ and $\eps'$, we have:

\[
\mathbb{E} \left[ \left| F(x,a,\eps)- F(x,a,\eps') \right|  \right]\leq [F]_L \mathbb{E} \left[\left| \eps- \eps' \right| \right]
\]

\vspace{2mm}

\noindent We now state the main result of this section whose proof is postponed in Appendix \ref{secappenquantif}.

\vspace{2mm}
\begin{theorem}
	\label{theo::Qknn}
	Take $K=M^{2+d}$ points for the optimal quantization of the exogenous noise $\eps_n$, $n=1,\ldots,N$. There exist constants $[\hat{V}^Q_n]_L>0$, that only depends on the Lipschitz coefficients of $f$, $g$ and $F$, such that, under \textbf{(Htrans)}, it holds for $n=0,...,N-1$, as $M\to +\infty$:
	\begin{equation}
	\label{ineq:theo2}
	\lVert \hat{V}^Q_n(X_n)- V_n(X_n) \rVert_2  \;\;\leq \;\;\sum_{k=n+1}^{N} \lVert r \rVert_\infty^{N-k}\left[\hat{V}_k^Q\right]_L \left( \eps_{k}^{proj} + [F]_L \eps_{k}^Q \right) + \mathcal{O}\left(\frac{1}{M^{1/d}}\right),
	\end{equation}
	where $\eps_{k}^Q:= \lVert \hat{\varepsilon}_k - \eps_k \rVert_2 $ stands for the quantization error, and $$\eps_{n}^{proj}:= \ssup{a \in A}\lVert  \mathrm{Proj}_{n+1}\left(F(X_n,a,\hat{\varepsilon}_{n})\right)  - F(X_n,a,\hat{\varepsilon}_{n})  \rVert_2$$ stands for the projection error, when decision $a$ is taken at time $n$.
\end{theorem}

\begin{remark}
	The constants $[\hat{V}^Q_n]_L>0$ are defined in \eqref{ineq:constlipVQ}.
\end{remark}

\vspace{3mm}

From Theorem \ref{theo::Qknn}, we can deduce consistency and provide a rate of convergence for the estimator $\hat{V}^Q_n$, $n=0,\ldots,N-1$, under some rather tough yet usual compactness conditions on the state space. 

\vspace{2mm}
\begin{corollary}
	\label{cor::Qknn}
	Under \textbf{(H$\mu$)} and \textbf{(HF)}, the Qknn-estimator $\hat{V}^Q_n$ is consistent for  $n=0,\ldots,N-1$, when taking $M^{d+1}$ points for the quantization; and moreover, we have for $n=0,...,N-1$, as $M \to +\infty$:
	\begin{equation}
	\lVert \hat{V}^Q_n(X_n)- V_n(X_n) \rVert_2 \;=\;  \mathcal{O} \left( \frac{1}{M^{1/d}} \right).
	\end{equation}
\end{corollary}

\vspace{3mm}
\begin{proof}
	We postpone the proof of Theorem \ref{theo::Qknn} to Appendix \ref{secappenquantif}.
\end{proof}

\subsection{Qknn agorithm applied to the order book control problem (\ref{eq1})}
\label{num}
\red{We recall the expression of the time-continuous controlled order book process \[Z_t= \big( X_t,Y_t,\underline{a}_t,\underline{b}_t,\underline{na}_t,\underline{nb}_t,pa_t,pb_t, \underline{ra}_t,\underline{rb}_t\big)\]  admits a representation as a MDP as shown in Section \ref{sec1}.
In Section \ref{secSol}, we proved that the value function associated to the MDP is characterized as limit as $N\to \infty$ of the the Bellman equation \eqref{eq:bellman1}. In this section, some implementation details on the Qknn algorithm are presented in order to numerically solve \eqref{eq:bellman1}.}

\noindent \textbf{Training set design}\vspace{2mm}\\
\noindent Inspired by  \citet{PaSa15}, we use product-quantization method and randomization techniques to build the training set $\Gamma_{n}$ on which we project $(T_n,Z_{n})$ that lies on $[0,T] \times E$, where $T_n$  and $Z_n$ stands for the $n$\textsuperscript{th} jump of $Z$ and the state of $Z$ at time $t_n$, i.e. $Z_n=Z_{T_n}$, for $n \geq 0$. This basic idea of Control Randomization consists in replacing in the dynamics of $Z$ the endogenous control by an exogenous control $(I_{T_{n}})_{n \geq 0}$, as introduced in \citet{lanetal14}. In order to alleviate the notations, we denote by $I_n$ the control taken at time $T_n$, for $n\geq 0$. \vspace{2mm}\\
\emph{Initialization.} Set: $\Gamma_0^E= \{z\}$ and $\Gamma^T_0= \{0\}$. \vspace{2mm}\\
Randomize the control, using e.g. uniform distribution on $A$ at each time step, and then simulate $D$ randomized processes to generate $(T_n^k,Z_n^k)_{n=0,k=1}^{N,D}$. \\

\vspace{-4mm}
\noindent For all $n=1,\ldots, N$, set $\Gamma^T_n=\{T_n^k, 1 \leq k \leq D\} $, which stands for the grid associated to the quantization of the $n^{th}$ jump time $T_n$, and set $\Gamma^E_n=\{Z_n^k,1 \leq  k \leq D\} $ which stands for the grid associated to the quantization of the state $Z_n$ of $Z$ at time $T_n$.
\begin{remark}
	The way we chose our training sets is often referred to as an \emph{exploration strategy} in the reinforcement learning literature. Of course, if one has ideas or good guess of where to optimally drive the controlled process, she should not follow an exploration-type strategy to build the training set, but should rather use the guess to build it, which is referred to as the \emph{exploitation strategy} in the reinforcement learning and the stochastic bandits literature. We refer to \citet{BalHurPha18} for several other applications of the \emph{exploration strategy} to build training sets. Note that this idea is the root of all the $Q$-learning based algorithms. See \citet{Sutton1998} for more details on $Q$-learning.
\end{remark}

\vspace{2mm}

Let $F$ and $G$ be the Borelian functions such that $ Z_n=F\big(Z_{n-1},d_n,I_n\big)$ and $ T_n=G\big(T_{n-1},\epsilon_n,I_n\big)$, where $\epsilon_n \sim \mathcal{E}(1)$ stands for the temporal noise, and $d_n$ is the state noise, for $n\geq 0$.

Let us fix $N \geq 1$ and consider $\big(\rwh{T}_n,\rwh{Z}_n\big)_{n=0}^N$, the dimension-wise projection of $\big(T_n,Z_n\big)_{n=0}^N$ on the grids $\Gamma^T_n \times \Gamma_n^E$, $n=0,\ldots,N$, i.e. $\rwh{T}_0 = 0$, $\rwh{Z}_0= z$, and 
\begin{equation}
\begin{cases}
\rwh{T}_n = \mathrm{Proj} \Big(G\big(\rwh{T}_{n-1}, \epsilon_n, I_n\big), \Gamma_n^T \Big), \\
\rwh{Z}_n=  \mathrm{Proj} \Big(F\big(\rwh{Z}_{n-1},d_n,I_n\big), \Gamma^E_n\Big), \qquad \text{ for } n= 1,\ldots,N.
\end{cases}
\end{equation}
$\big(\rwh{T}_n,\rwh{Z}_n,I_n\big)_{n \in \{0,N\}}$ is a Markov chain. 

Define then $\big(\rwh{T}_n^Q,\rwh{Z}_n^Q\big)_{n=0}^N$ as temporal noise-quantized version of $\big(\rwh{T}_n,\rwh{Z}_n,I_n\big)_{n=0}^N$. Note that we do not need to quantize the spacial noise since this noise already takes a finite number of states. Let $\hat{\varepsilon}_n$ be the quantized process associated to $\epsilon_n$. The process  $\big(\rwh{T}_n^Q,\rwh{Z}_n^Q\big)_{n=0}^N $ is then defined as follows: $\rwh{Z}^Q_0= z$, $\rwh{T}_0^Q=0$ and $ \forall 1 \leq n \leq N$:
\begin{equation}
\begin{cases}
\rwh{T}^Q_n = \mathrm{Proj} \Big(G\big(\hat{t}_{n-1}, \hat{\varepsilon}_n, I_n\big), \Gamma_n^T \Big), \\
\rwh{Z}^Q_n= \mathrm{Proj}\Big(F\big(\rwh{Z}_{n-1},d_n,I_n\big), \Gamma^E_n\Big).
\end{cases}
\end{equation}

\noindent Denote by $\left(\rwh{V}^{Q,(N,D)}_n\right)_{n=0}^N$ the solution of the Bellman equation associated to $\left(\rwh{T}_n^Q,\rwh{Z}_n^Q\right)_{n=0}^N$: 
\begin{equation}
(\rwh{B}^Q_{N,D}):	\left\{
\begin{array}{ccl}	
\rwh{V}^{Q,(N,D)}_N &=&0 \\
\rwh{V}^{Q,(N,D)}_{n}(t,z)&=&r(t,z,a)  + \ssup{a \in A} \left\{ \mathbb{E}_{t,z}^a\left[\rwh{V}^{Q,(N,D)}_{n+1}\big(\rwh{T}^Q_{n+1},\rwh{Z}^Q_{n+1}\big) \right] \right\}, \text{ for } n=0, \ldots,N,
\end{array}
\right.
\end{equation}
where $\E_{t,z}^a[.]$ stands for the expectation conditioned by the events $ \rwh{T}^Q_n=t$,$\rwh{Z}^Q_n=z$ and when decision $I_n=a$ is taken at time $t$.
\vspace{4mm}\\
We wrote the pseudo-code of the Qknn algorithm to compute $(\rwh{B}^Q_{N,D})$ in Algorithm \ref{algo:Qknn_LOB}.\\

\begin{algorithm}
	\caption{Generic Qknn Algorithm}
	\label{algo:Qknn_LOB}
	\textbf{Inputs}:\\ 
	-- $N$: number of time steps\\
	-- $z$: state in $E$ at time $T_0=0$\\
	-- $\Gamma^\eps= \{e_1,\ldots,e_L\}$ and $(p_\ell)_{\ell=1}^L$: the grid and the weights for the optimal quantization of $(\eps_n)_{n=1}^N$.\\
	-- $\Gamma_{n}$ and $\Gamma_n^E$ the grids for the projection of respectively the time and the state components at time $n$, for $n=0,\ldots,N$. 
	\begin{algorithmic}[1]
		\For{ $n$ $=$ $N-1,\ldots,0$}
		\State  Compute the approximated $Qknn$-value at time $n$:
		\beq
		\hat Q_n(z,a) & = &  r\left(T_n,z,a \right)\\
		&& +\sum_{\ell=1}^L p_\ell \widehat{V}_{n+1}^Q\left( \text{Proj} \left(  G(z,e_\ell,a), \Gamma_{n+1}^T \right), \text{Proj} \left(  F(z,e_\ell,a), \Gamma_{n+1}^E \right) \right),\\
		&& \hspace{7cm}\text{ for } (z,a) \in \Gamma_{n} \times A_z; \nonumber 
		\enq
		\State Compute the optimal control at time $n$ 
		\begin{equation*}
		\hat A_n(z) \;\; \in\;\; \argmin_{a \in A_z} \;\hat Q_n(z,a), \quad \text{ for } z \in \Gamma_{n},
		\end{equation*}
		where the $\argmin_{}$ is easy to compute since $A_z$ is finite for all $z \in E$;
		\State  Estimate analytically by quantization the value function: 
		\begin{equation*}
		\widehat{V}_{n}^Q(z) \;\; = \; \;  \hat Q_n \left(z,\hat A_n(z) \right), \quad \;\;\; \forall z \in \Gamma_n;
		\end{equation*}
		\EndFor
	\end{algorithmic}
	\textbf{Output:} \\
	-- $(\widehat{V}_{0}^Q)$: Estimate of $V(0,z)$;\\
\end{algorithm}

We discuss in Remark \ref{borneZ} the reasons why we can apply Theorem \ref{theo::Qknn}.
\begin{remark}
	\label{borneZ}
	When the number of jumps of the LOB $N\geq 1$ is fixed, the set of all the states that can take the controlled order book by jumping less than $N$ times, denoted by $\mathcal{K}$ in the sequel, is finite. 
	Hence, the reward function $r$, defined in \eqref{eq:reward},  is bounded and Lipschitz on $\mathcal{K}$.
\end{remark}

The following  proposition  states that $\rwh{V}^{Q,(N,D)}_n$, built from the combination of time-discretization, $k$-nearest neighbors and optimal quantization methods, is a consistent estimator of the value function at time $T_n$, for $n=0,\ldots,N-1$. It provides a rate of convergence for the Qknn-estimations of the value functions.

\begin{proposition}
	\label{cor::Qknnconsistant}
	The estimators of the value functions provided by Qknn algorithm are consistent. Moreover, it holds as $M \to +\infty$:
	\begin{equation}
	\left \lVert  \rwh{V}^{Q,(N,M)}_n\left(\rwh{T}_n,\rwh{Z}_n\right)- V_n\big(T_n,Z_n\big)  \right\rVert_{M,2} = \mathcal{O} \left(\alpha^N + \frac{1}{M^{2/d}} \right), \quad \text{for } n=0,\ldots,N-1, 
	\end{equation}
	where we denote by $\lVert. \rVert_{M,2}$ the $\mathbb{L}^2(\mu)$ norm conditioned by the training sets that have been used to build the estimator $ \rwh{V}^{Q,(N,M)}_{n+1}$.
\end{proposition}

\begin{proof} 
	Splitting the error of time cutting and quantization, we get: 
	\begin{align}
	\lVert V_n\big(T_n,Z_n\big) -  \rwh{V}^{(N,M)}_n\big(\rwh{T}_n,\rwh{Z}_n\big) \rVert_{M,2} & \leq\lVert V^{}_n\big(T_n,Z_n\big) -  V^{(N)}_n\big(T_n,Z_n\big) \rVert_{M,2} \\
	&\hspace{.5cm}+ \lVert V^{(N)}_n\big(T_n,Z_n\big) -  \rwh{V}^{(N,M)}_n\big(\rwh{T}_n,\rwh{Z}_n\big) \rVert_{M,2}. \label{ineq:splitError}
	\end{align}
	\emph{Step 1:} Applying Lemma \ref{lem::cuthorizon}, we get the following bound on the first term in the r.h.s. of \eqref{ineq:splitError}:
	\begin{equation}
	\label{ineq:ineq1}
	\lVert V_n\big(T_n,Z_n\big) -  V^{(N)}_n\big(T_n,Z_n\big) \rVert_{M,2}  \leq \frac{\alpha^N}{1-\alpha } \normeinf{b},
	\end{equation}
	where $\normeinf{b}$ stands for the supremum of $b$ over $[0,T] \times E$.\vspace{2mm}\\
	\emph{Step 2:} Note that the assumptions of Theorem \ref{theo::Qknn} are met as noticed in Remark \ref{borneZ}, so that the latter provides the following bound for the second term in the r.h.s. of \eqref{ineq:splitError}:
	\begin{equation}
	\label{ineq:ineq2}
	\left \lVert  V^{(N)}_n\big(T_n,Z_n\big) -  \rwh{V}^{Q,(N,M)}_n\big(\rwh{T}_n,\rwh{Z}_n\big) \right\rVert_{M,2}\underset{M \to \infty}{=} \mathcal{O} \left( \frac{1}{M^{2/d}} \right).
	\end{equation}
	
	\noindent It remains to plug \eqref{ineq:ineq1} and \eqref{ineq:ineq2} into \eqref{ineq:splitError} to complete the proof of Proposition \ref{cor::Qknnconsistant}.
\end{proof}

\subsection{Numerical results}
\label{sec:LOBnumres}
In this section, we propose several settings to test the efficiency of Qknn on simulated order books.  We take no running reward, i.e. $f=0$, and take the wealth of the market maker \red{after liquidating their inventory} as terminal reward, i.e. \red{$g(z)=x + L(y)$}. The intensities are taken constant in the first tests, and state dependent in the second ones. The values of the state dependent intensities are similar to the ones in \citet{RoLe} (see ).
\begin{figure}
	\centering
	\includegraphics[width=1\linewidth]{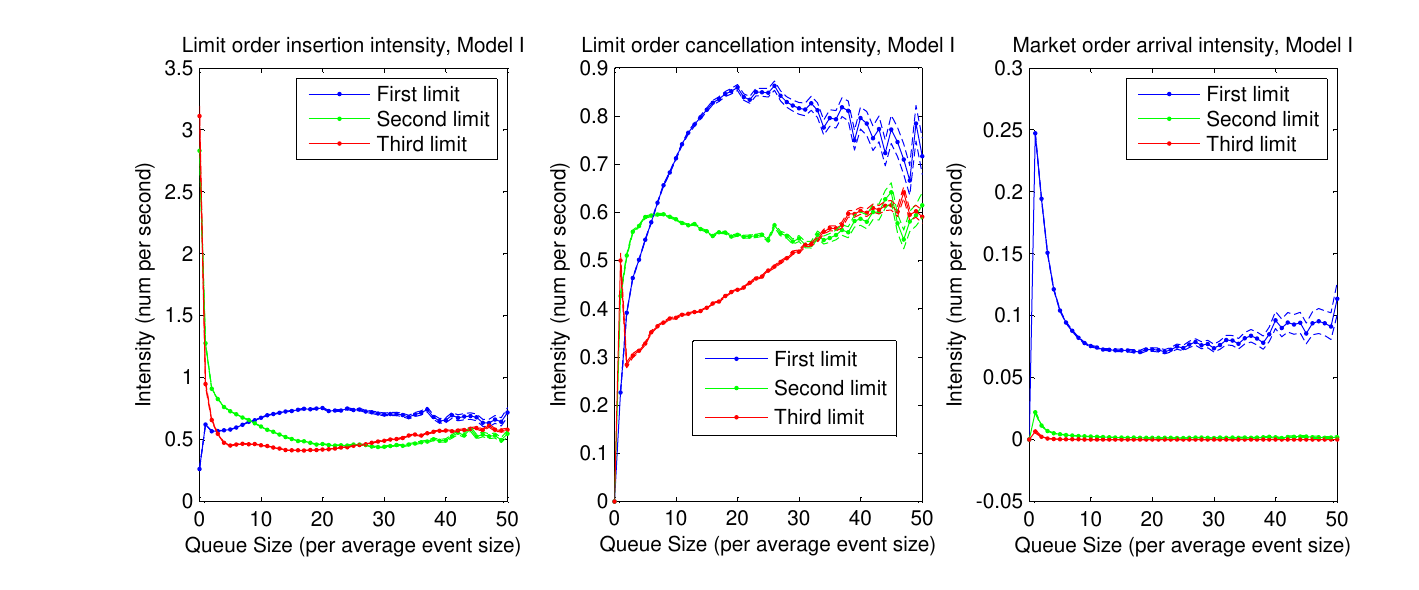}
	\caption{\red{Values of the intensities w.r.t. the queue size used in the numerical tests of Section \ref{sec:LOBnumres}. The plot is taken from \citet{RoLe} (see its Figure 13). The intensities are estimated by the authors using their data from Alcatel Lucent. }}
	\label{fig:intensitymodel1}
\end{figure}
 Although the intensities are assumed to be uncontrolled in section \ref{existenceCharacterisation} for predictability reasons, the latter are controlled processes in this section, i.e. the intensities of the order arrivals depends on the orders in the order book from all the participant plus the ones of the market maker. The optimal trading strategies have been computed among two different classes of strategies: in Section \ref{case1}, we tested the algorithm to approximate the optimal strategy among those where the market maker is only allowed to place orders only at the best bid and the best ask; in Section \ref{sec2}, we computed the optimal trading strategy among the class of the strategies where the market maker allows herself to place orders on the two best limits on each side of the order book. Note that the second class of controls is more general than the first one. The code is available on \url{https://github.com/comeh}.

The search of the $k$ nearest neighbors, that arises when estimating the conditional expectations using the Qknn algorithm, is very time-consuming; especially in the considered market-making problem which is of dimension more than 10. The efficiency of Qknn then highly depends on the algorithm used to find the $k$ nearest neighbors in high-dimension. We implemented the method using the Fast Library for Approximate Nearest Neighbors algorithm (FLANN), introduced in \citet{MuLo} and already available as a library of C++, Python, Julia and many other languages. This algorithm is based on tree methods. Note that recent algorithms based on graph also proved to perform well and can also be used.  

\subsubsection{Case 1: The market maker only place orders at the best ask and best bid}
\label{case1}
Denote by A1lim the class of controls where the placements of \red{orders are allowed} on the best ask and best bid exclusively. We implement the Qknn algorithm to compute the optimal strategy among those in A1lim. We then compared the optimal strategy with a naive strategy which consists in always placing one order at the best bid and one order at the best ask. The naive strategy is called 11 in the plots, and can be seen as a benchmark. The naive strategy is a good benchmark when the model for the intensities of order arrivals is symmetrical, i.e. the intensities for the bid and the ask sides are the same. Indeed, in this case, the market maker can expect to earn the spread in average. \\

\begin{figure}
	\centering
		\begin{subfigure}{.5\linewidth}
	\includegraphics[width=1.2\linewidth]{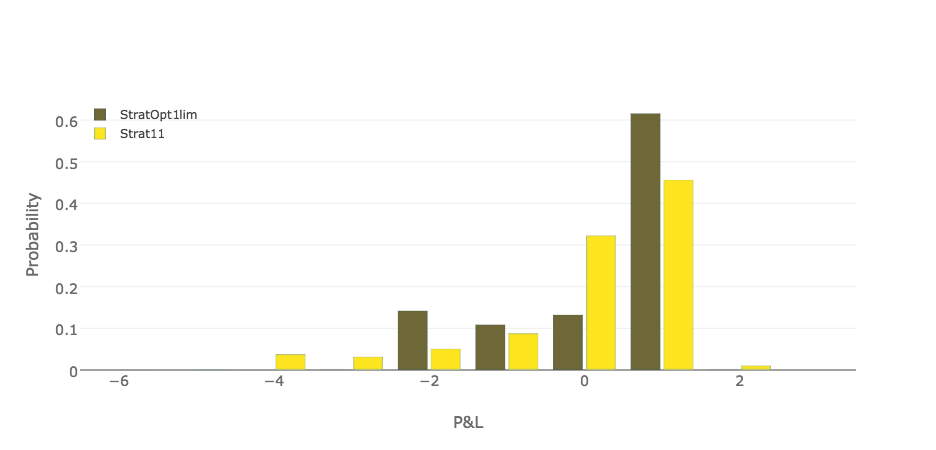}
\caption{$10^5$ points for the grids.}
\label{figConst}
	\end{subfigure}
	\begin{subfigure}{.45\linewidth}
	\centering 
\includegraphics[width=9cm, trim=0 0 0 100, clip=true]{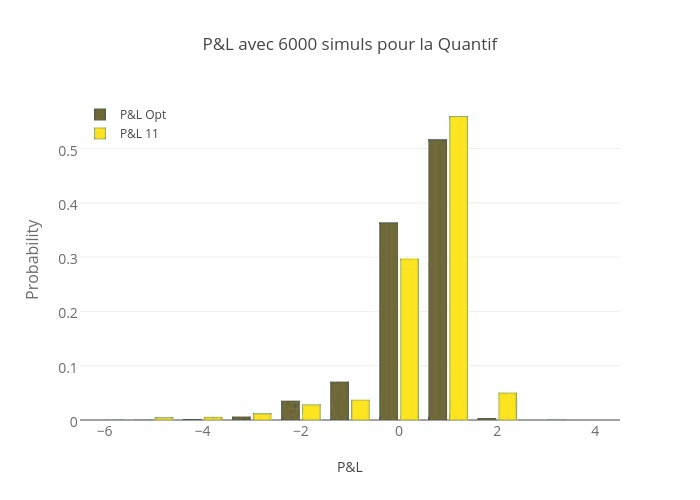}
\label{fig:sub2}
\vspace{-7.5mm}
\caption{6000 points for the grids.}
\label{fig:test}
\end{subfigure}
	\caption{Histogram of the P\&L when following the Qknn estimated optimal strategy (Opt) and the naive strategy (11). We took symmetrical and constant intensities, and a short terminal time $T=1$. Notice that the Qknn strategy looks to improve the P\&L by reducing the losses when enough points are taken to build the grids (see Figure \ref{figConst}), and that its performance is worse if less points are taken to build the grids (see Figure \ref{fig:test}).}
\end{figure}

In Figure \ref{figConst}, we take constant intensities to model the limit and market orders arrivals, and linear intensity to model the cancel orders. In this setting, as we can see in the figure, the strategy computed using Qknn algorithm performs as well as the naive strategy. Note that, obviously, the market maker has to take enough points for the state quantization in order for Qknn algorithm to perform well. In Figure \ref{fig:test}, we plotted the P\&L of the market maker when the latter compute the optimal strategy using only 6000 points for the state space discretization, and for such a low number of points for the grid, Qknn algorithm performs poorly. \red{In this setting, notice that the naive strategy seems to perform well.}

\begin{figure}%
	\centering
	\begin{subfigure}{.5\linewidth}
		\includegraphics[width=8cm, trim=0 0 0 100, clip=true]{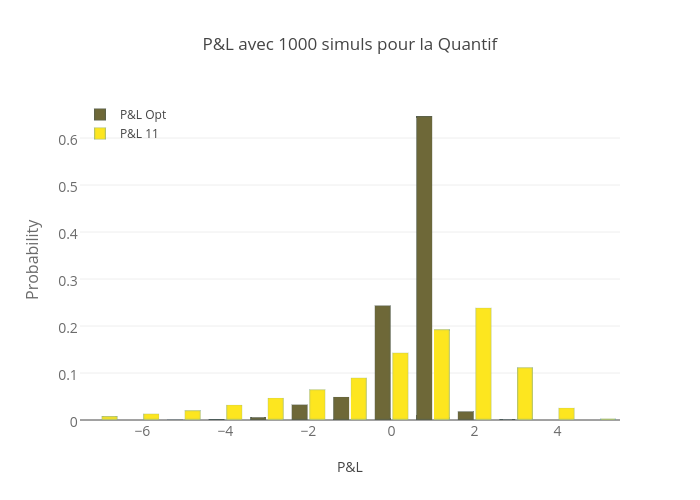} 
		\caption{(a)}
		\label{1000ptsquantif} %
	\end{subfigure}
	\begin{subfigure}{.4\linewidth}
		\includegraphics[width=8cm, trim=0 0 0 100, clip=true]{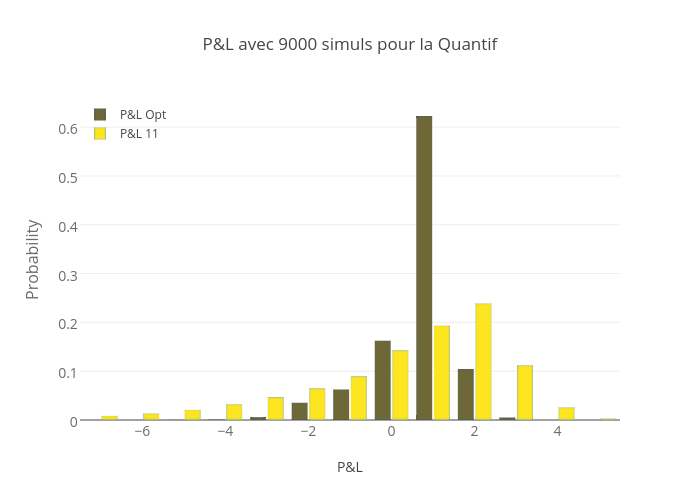}
		\caption{(b)}
		\label{9000ptsquantif}
	\end{subfigure}
	\begin{subfigure}{.5\linewidth}
		\includegraphics[width=8cm, trim=0 0 0 100, clip=true]{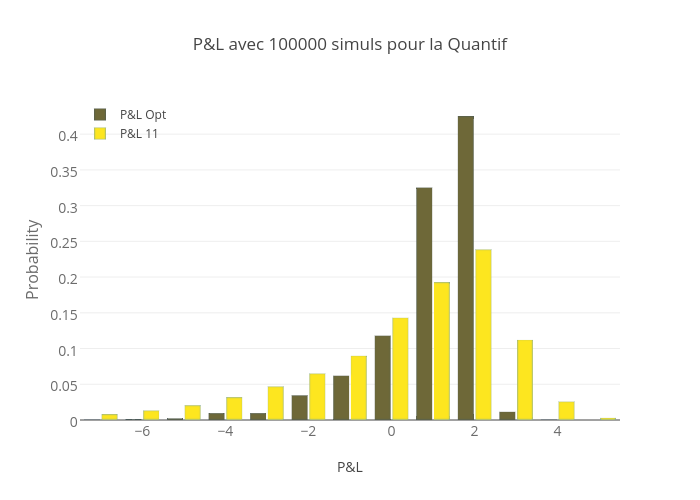}
		\caption{(c)}
		\label{100000ptsquantif}
	\end{subfigure}
	\begin{subfigure}{.4\linewidth}
		\includegraphics[width=8cm, trim=0 0 0 100, clip=true]{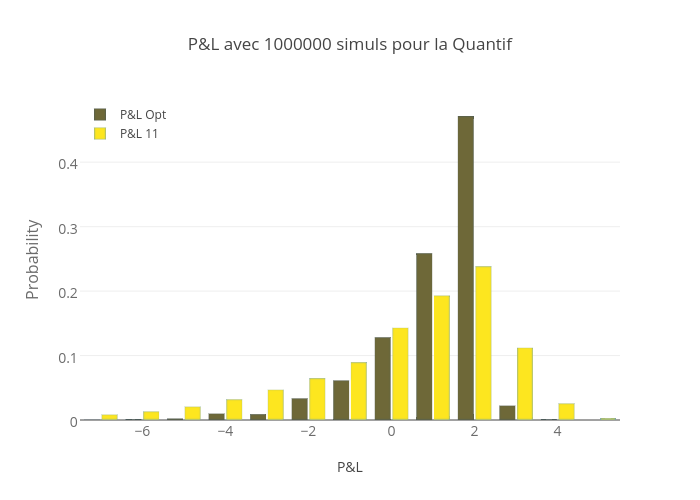}
		\caption{(d)}
		\label{1000000ptsquantif}
	\end{subfigure}
	\caption{Histogram of the P\&L when the market maker follow the Qknn estimated optimal strategy (Opt) and the naive strategy (11). The intensities $\lambda^M$ , $\lambda^L_i$ and $\lambda^C_i$ depend on the state of the order book. The P\&L of the market maker when following the Qknn-estimated optimal strategy is computed using $10^3$, $10^4$, $10^5$, and $10^6$ points for the grids: see respectively Figures \protect\ref{1000ptsquantif}, \protect\ref{9000ptsquantif}, \protect\ref{100000ptsquantif}  and \protect\ref{1000000ptsquantif}. The reader can see that the market maker increases their expected terminal wealth (after liquidation) by taking more and more points for the state space discretization. Also, the naive strategy is beaten when the intensities are state dependent. }%
	\label{figdep}%
\end{figure}

In Figure \ref{figdep}, we plotted the empirical histogram of the P\&L of the market maker using the Qknn-estimated optimal strategy, computed with grids of size $N=10^3,10^4,10^5, 10^6$ for the state space discretization; and the empirical histogram of the P\&L of the market maker using the naive strategy. We took intensities that are state dependent.
One can see that the larger the size of the grids are, the better the Qknn-estimation of the optimal strategy is.

\begin{figure}
	\centering
	\includegraphics[width=.7\linewidth,keepaspectratio=true,trim={0 .5cm 0 5cm},clip ]{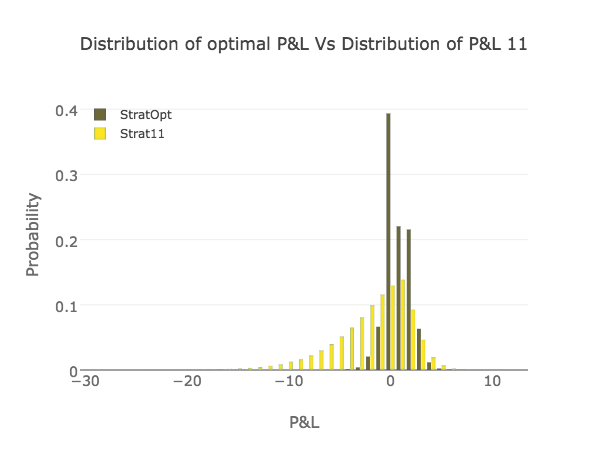}
	\caption{Distribution of the P\&L when the market maker follows the optimal strategy (dark blue) and the naive strategy (light blue). We took symmetrical and state dependent intensities; and long terminal time: $T=10$.  Notice that the Qknn strategy does better than the naive strategy when the intensities are state dependent.}
	\label{fig:Solution1}
\end{figure}

In Figure \ref{fig:Solution1}, we plot the P\&L of the market maker following the Qknn-estimated optimal strategy and the naive strategy. We took the same parameters as in Figure \ref{figdep} to run the simulations except from the terminal time that we set to be equal to $T=10$. In this setting, the Qknn-estimated optimal strategy performs much better than the naive strategy, which highlights the fact that the naive strategy is not optimal. 

\captionsetup[subfigure]{labelformat=empty}

\begin{figure}
	\centering
	\includegraphics[width=.7\linewidth,keepaspectratio=true,trim={0 .5cm 0 5cm},clip ]{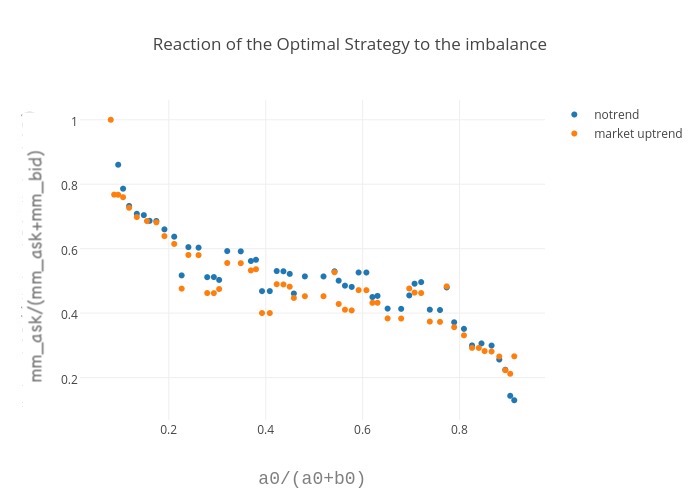}
	\caption{Reaction of Qknn to an artificial positive trend in the LOB. The y-axis represents the ratio of orders sent on the ask side by the market maker to their orders sent in the ask and bid sides in the order book. The x-axis is represents the ratio of the size of the best-ask limit to the sum of the sizes of the best-bid and best-ask limits. The blue points are those without trend in the market. The orange points are those with a positive trend in the market. We can see that Qknn took correctly the trend into account in its decisions: for two same order books, Qknn is less willing to sell when the price is expected to increase.
	}
	\label{fig:NotrendVsTrend}
\end{figure}

\red{We plotted in Figure \ref{fig:NotrendVsTrend} the reaction of Qknn when a trend is added in the dynamic of the market. In this example, we took a higher intensity for the sell market order than the one for the buy market order, which creates an artificial positive trend in the dynamic of the price.  Observe that Qknn understood correctly that it is better not to sell when the price goes up.}

\subsubsection{Case 2: the market maker place orders on the first two limits of the Orders Book}
\label{sec2}
\begin{figure}
	\centering
	\includegraphics[width=.7\linewidth,keepaspectratio=true,trim={0 1cm 0 3.5cm},clip ]{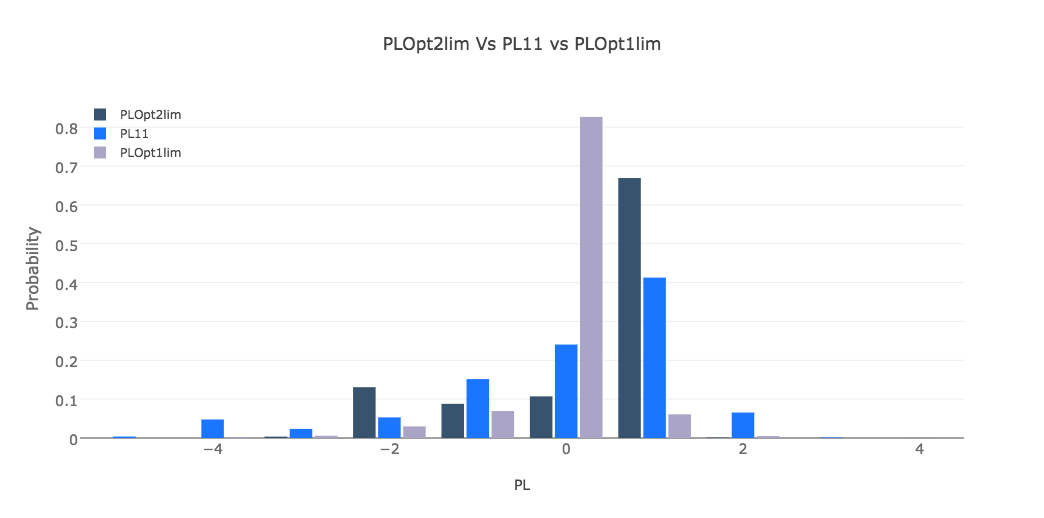}
	\caption{P\&L of the market maker who follows optimal strategies and the naive strategy (PL11). Short Terminal Time. asymmetrical intensities for the market order arrivals: the intensity for the buying market order process is taken higher than the one for the selling market order process. The wealth of the market maker is greater when she places orders on the two first limits of each sides of the order book, rather than when she places orders only on the best limits at the bid and ask sides. 
	}
	\label{fig:2limits_1}
\end{figure}

\begin{figure}
	\centering
	\includegraphics[width=1.1\linewidth,keepaspectratio=true,trim={0 .5cm 0  3cm},clip]{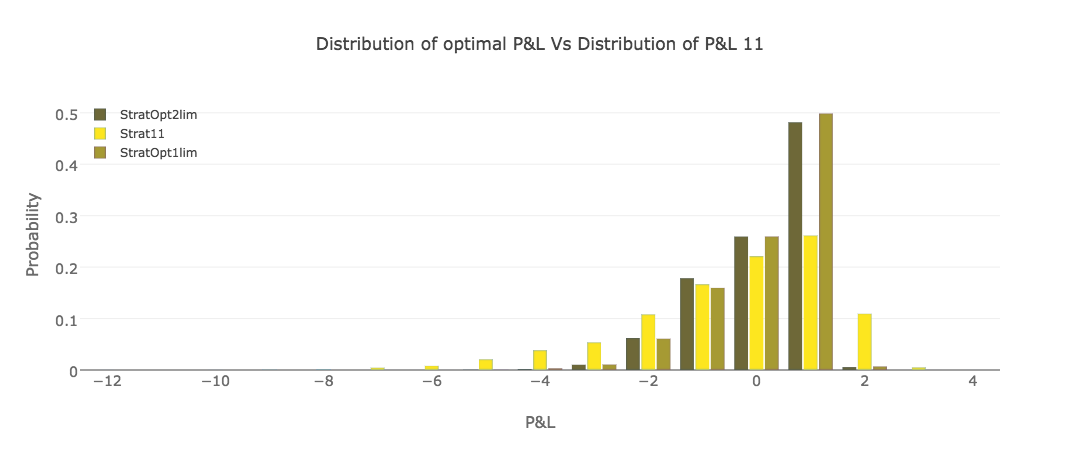}
	\caption{P\&L when following the optimal strategy or the naive strategy (PL11). Long Terminal Time. Symmetrical intensities for the arrival of market orders.  $4.10^4$ points for the quantization. Notice that the Qknn strategy computed on the extended class of controls, i.e. order placements on the two first limits (StratOpt2lim),  performs as well as the one computed on the original class of controls, i.e. order placements on the best-bid and best-ask (StratOpt1lim). 
	}
	\label{fig:2limits_2}
\end{figure}

We extend the class of admissible controls to the ones where the market maker places order on the first two limits on the bid and ask sides of the order book. Denote by $A2lim$ the latter. We run simulations to test the Qknn algorithm on A2lim.
In figure \ref{fig:2limits_1} and figure \ref{fig:2limits_2}, we plot the empirical distributions of the  P\&L  when the market maker follows the three different strategies: 
\begin{itemize}
	\item {\color[RGB]{55, 83, 109} Qknn-estimated optimal strategy among those in A2lim (PLOpt2lim).}
	\item {\color{CadetBlue} Qknn-estimated optimal strategy among those in A1lim (PLOpt1lim).}
	\item {\color[RGB]{26, 118, 255}  naive strategy, i.e. always place orders on the best bid and best ask queues (PL11).}
\end{itemize}
Note that the P\&L of the market maker is always better when the class of admissible controls is extended, see Figure \ref{fig:2limits_1}; but in some numerical tests, the extended set of controls does not seem to improve the P\&L. Indeed, we observed that the terminal P\&L estimated using Qknn among A2lim and A1lim have the similar empirical distribution in the tests whose results are presented in Figure \ref{fig:2limits_2}.

\newpage
\section{Model extension to Hawkes Processes}
\label{sec::hawkes}

We consider in this section a market maker who aims at maximizing a function of their terminal wealth, penalizing their inventory at terminal time $T$ in the case where the orders arrivals are driven by Hawkes processes. Let us first present the model with Hawkes processes for the LOB.

\noindent \textbf{Model for the LOB:} 

We assume that the order book receives limit, cancel, and market orders. We denote by $L^{+}$ (resp. $L^{-}$) the limit order arrivals process the ask (resp. bid) side; by $C^{+}$ (resp. $C^{-}$) the cancel order on the ask (resp. bid) side; and by $M^{+}$ (resp. $M^{-}$) the buy (resp. sell) market order arrivals processes.  In this section, the limit orders arrivals are  assumed to follow Hawkes processes dynamics, and moreover we assume the kernel to be exponential. The order arrivals are then modeled by a (4K+2)-variate Hawkes process $(N_t)$ with a vector of exogenous intensities $\lambda_0$ and exponential kernel $\phi$, i.e. $\phi^{ij}(t)=\alpha_{ij}\beta e^{\beta t} \1_{t \geq 0}$. 

\red{Let $\alpha$=$(\alpha^{ij})_{i,j }$.
We assume: $(\alpha)_{i,j}$ to have spectral radius strictly smaller than 1, which is a sufficient condition to guarantee stationarity of the model and convergence of $\E[\lambda_u | \mathcal{F}_t ]$ as $u\to +\infty$, as shown e.g. in \cite{hawkes1971spectra}. } 

Note that in the presented model, the following holds:\vspace{2mm}\\
\noindent \textbf{(H$\lambda$)} $\lambda$ is assumed to be independent of the control.
\vspace{2mm}\\
Denoting by $D=4K+2$ the dimension of $(N_t)$, the $m$\ts{th} component of the intensity $\lambda$ of $N_t$ writes, under \textbf{(H$\lambda$)}: 
\begin{equation}
\lambda^m_t= \lambda_0^m + \sum_{j=1}^{D}\alpha_{mj} \int_0^te^{-\beta(t-s)} \diff N^j_s, \qquad \qquad \text{ for } m=1,\ldots, D,
\end{equation} 
It is well-known that for this choice of intensity, the couple $(N_t, \lambda_t)_{t\geq 0}$ becomes Markovian, see e.g. Lemma 6 in \citet{Ma98} for a proof of this result, and moreover we have: 
\begin{equation}
\diff \lambda^m_t= -\beta D \big( \lambda_t^m -\lambda_0^m\big) dt + \sum_{j=1}^{D} \alpha_{mj} \diff N^j_t, \qquad \qquad \text{ for } m=1,\ldots, D,
\end{equation}
with given initial conditions: $ \lambda_0^m \in \R_+^*$ for $m=1,\ldots, D$.

We can now rewrite the control problem \eqref{eq1} in the particular case where the order book is driven by Hawkes processes, there is no running reward, i.e. $f=0$, and where the terminal reward $G$ stands for the terminal wealth of the market maker penalized by their inventory. We then consider the following problem in this section:
\begin{equation}
\label{eq2}
V(t,\lambda,z) := \ssup{\alpha \in \mathbb{A}} \mathbb{E}_{t,z,\lambda}^\alpha \big[ G\big(Z_T \big)\big],
\end{equation}
where $G(z)$ denotes the wealth of the market maker when the controlled order book is at state $z$, plus a term of penalization of their inventory; and where $\mathbb{A}$ is the set of the admissible controls, i.e. the predictable decisions taken by the market maker until a terminal time $T >0$.\\

\noindent We now present the main result of this section.

\begin{theorem}
	\label{thm:hk}
	$V$ is characterized as the unique solution of the following HJB equation:
	\begin{equation}
	\label{eq:car1HK}
	\left\{
	\begin{array}{rcl}
	f(T,z,\lambda) &=& G(z), \text{ for } z \in E\\
	0&=& \displaystyle \frac{\partial f}{\partial t}(t,z,\lambda)  -  \displaystyle D\beta \sum_{m=1}^{D} \left[\big(\lambda^m - \lambda^m_0\big) \frac{\partial f}{\partial \lambda^m}(t,z,\lambda) \right. \\
	&& \hspace{2cm}  \left. +  \lambda^m  \ssup{a \in A_z} \big[ f\big(t,e_m^a(z), \lambda + \alpha_m\big) - f\big(t,z, \lambda \big)\big] \right], \\
	&& \hspace{5cm}\text{ for } 0 \leq t < T, \; \text{ and } \; (t,z,\lambda) \in \R_+ \times E \times \R_+^*.
	\end{array}
	\right.
	\end{equation}
	where $\alpha_m=(\alpha_{1m}, \ldots, \alpha_{Dm})$.
	Moreover, $V$ admits the following representation:
	\begin{align}
	V(t,z,\lambda)  &=\ssup{\alpha \in \mathbb{A}}  \sum_{n=0}^{\infty } \mathbb{E}^\alpha_{t,z,\lambda} \left[ 1_{T_n \leq T} G\left(  Z_{T_n}^\alpha \right) \exp\Bigg\{ -\abs{\lambda_0}(T-T_n)  \right.\\
	&\hspace{3.5cm} + \left. \left. \sum_{m=1}^{D} \frac{\lambda^m_{T_n}- \lambda_0^m}{D\beta}\left( e^{-\sum_{j=1}^{D}\beta_{mj}(T-T_n)}-1\right) \right\}\right], \label{eq::valuefunctionPDMDP_1}
	\end{align}
	where, for $n\geq 0$, $T_n$ stands for the $n$\ts{th} jump time of $Z$ after time $t$, and  $(Z_{T_{n}}^{\alpha},\lambda_{T_n})_{n=0}^\infty$ is as a MDP controlled by $\alpha \in \mathbb{A}$; and where $\mathbb{E}_{t,z,\lambda}^\alpha[.]$ stands for the expectation conditioned by $Z_t=z, \lambda_t=\lambda$ when the control $\alpha$ is followed.
\end{theorem}
\vspace{2mm}
\begin{remark}
	$V$ is characterized in  \eqref{eq::valuefunctionPDMDP_1} as the value function associated to an MDP with infinite horizon, where the running reward reads:
	\begin{equation}
	\begin{array}{rcl}
	r(t,z,\lambda)\;&=&\; \mathds{1}_{t \leq T} \;G\left( z \right) \exp\left\{ -\abs{\lambda_0}_1(T-t) + \displaystyle \sum_{m=1}^{D} \frac{\lambda^m - \lambda_0^m}{D\beta}\left( e^{-D\beta(T-t)}-1\right) \right\},
	\end{array}
	\end{equation}
	where $\vert .\vert_1$ denotes the $\mathbb{L}^1 \big(\R^D \big)$ norm.
\end{remark}
\vspace{2mm}
\noindent \emph{Proof:} (of Theorem \ref{thm:hk}) \vspace{2mm}\\
\noindent \emph{Step 1:} Let us check that \eqref{eq::valuefunctionPDMDP_1} holds, where $V$ is defined as solution of \eqref{eq2}.\\

 First notice that $(\lambda_t,Z_t)_t$ is a PDMDP\footnote{PDMDP stands for Piecewise Deterministic Markov Decision Process, which is a MDP whose dynamic is deterministic between jumps.}, i.e. $(\lambda_t,Z_t)_t$ is deterministic between two jumping times. We then aim at rewriting the expression of the value function defined in \eqref{eq2} as the value function associated to a infinite horizon control problem of the PDMDP $(\lambda_t,Z_t)_t$. To do so, we first notice that by conditioning on the time jumps we get:
\begin{align}
V(t,z,\lambda) &= \ssup{\alpha \in \mathbb{A}}  \mathbb{E}_{t,z,\lambda}^\alpha \Big[ G \big(Z_T^\alpha \big) \Big] \nonumber \\
&= \ssup{\alpha \in \mathbb{A}} \mathbb{E}_{t,z,\lambda}^\alpha \bigg[ \sum_{n=0}^{\infty } 1_{T_n \leq T < T_{n+1}} G\big(  Z_{T_n}^\alpha \big) \bigg] \nonumber \\
&= \ssup{\alpha \in \mathbb{A}}  \sum_{n=0}^{\infty } \mathbb{E}_{t,z,\lambda}^\alpha  \bigg[ 1_{T_n \leq T} G\big(  Z_{T_n}^\alpha \big) \mathbb{P}\Big(T-T_n \leq T_{n+1}-T_n \big\vert T_n \Big) \bigg], \label{eqpd}
\end{align}
where  $\big(T_n\big)_n$ is the sequence of jump times of $N$. 
This process is a jump process with intensity $\mu_s=\sum_{m=1}^{D} \lambda^m_s$.
Since it holds, conditioned on $\mathcal{F}_{T_n}$:  
\[\mu_s= \sum_{m=1}^{D}\lambda_0^m+ \big( \lambda^m_{T_n}- \lambda^m_0 \big)e^{-{D} \beta(s-T_n)}, \qquad  \text{ for }  s \in [T_n, T_{n+1}), \]
then, we have: 
\begin{align}
\mathbb{P}\Big(T_{n+1}-T_n \geq T-T_n \big\vert T_n\Big) &= \int_{T-T_n}^{\infty} \mu_s e^{-\int_{0}^{s} \mu_udu} \diff s \\
& \hspace{-3cm}= \exp\Bigg\{ -\abs{\lambda_0}(T-T_n) + \sum_{m=1}^{D} \frac{\lambda^m_{T_n}- \lambda_0^m}{D\beta}\Big( e^{-D\beta(T-T_n)}-1\Big) \Bigg\}. \label{eq::proba}
\end{align}
Plugging \eqref{eq::proba} into \eqref{eqpd}, the value function rewrites:
\begin{align}
V(t,z,\lambda)  &=\ssup{\alpha \in \mathbb{A}}  \sum_{n=0}^{\infty } \mathbb{E}^\alpha_{t,z,\lambda} \left[ 1_{T_n \leq T} G\left(  Z_{T_n}^\alpha \right) \exp\Bigg\{ -\abs{\lambda_0}(T-T_n)  \right.\\
&\hspace{3.5cm} + \left. \left. \sum_{m=1}^{D} \frac{\lambda^m_{T_n}- \lambda_0^m}{D\beta}\left( e^{-D\beta(T-T_n)}-1\right) \right\}\right], \label{eq::valuefunctionPDMDP}
\end{align}
which completes the step 1.
The r.h.s of \eqref{eq::valuefunctionPDMDP} can be seen as the value function of an infinite horizon control problem associated to the PDMDP.
\vspace{3mm}

\noindent \emph{Step 2:} Let us show that $V$ is the unique solution to \eqref{eq:car1HK}. 

\noindent Notice first that the solutions to the following HJB equation 
\begin{equation}
\left\{
\begin{array}{rcl}
G(z) &=&f(T,z,\lambda) \\
0&=&\frac{\partial f}{\partial t}  - \sum_{m=1}^{D}  D \beta  \big(\lambda^m - \lambda^m_0\big) \frac{\partial f}{\partial \lambda^m} +  \lambda^m  \ssup{a \in A_z} \big[ f\big(t,e_m^a(z), \lambda+ \alpha_m\big) - f\big(t,z, \lambda \big)\big], \\
&& \hspace{8cm}\text{ for } 0 \leq t < T.
\end{array}
\right.
\end{equation}
are the fixed points of the operator $\mathcal{T}= \mathcal{T}_1 \circ \mathcal{T}_2$ where $ \mathcal{T}_1$ and $\mathcal{T}_2$ are defined as follows: 
\[
\mathcal{T}_1 : F \mapsto f \text{ solution of } 
\begin{cases}
\frac{\partial f}{\partial t}  - D \beta \sum_{m=1}^{D}  \big(\lambda^m - \lambda^m_0\big) \frac{\partial f}{\partial \lambda^m} = F(t,z,\lambda) \\
f(T,z,\lambda) = G(z),
\end{cases}
\]
and: 
\[
\mathcal{T}_2: f \mapsto - \sum_{m=1}^{D} \lambda^m  \ssup{a \in A_z} \big[ f\big(t,e_m^a(z), \lambda+ \alpha_m\big) - f\big(t,z, \lambda \big)\big].
\]

We now use the characteristic method to rewrite the image of $\mathcal{T}_1$. \\Let us take function $F$, and define $f= \mathcal{T}_1(F)$.
Let us fix $t \in [0,T]$ and $\lambda \in (\mathbb{R}_+)^D$, and denote by $g$ the function $g(s,z)=f(s,z,\lambda^1_s, ..., \lambda^D_s)$ where, for $m=1,\ldots,D$, $s \mapsto \lambda^m_s$ is a  differentiable function defined on $[t,T]$ as solution to the following ODE:
\begin{equation}
\left\{
\begin{array}{rcl}
\frac{d \lambda_s^m }{ds} &=& -D\beta\big(\lambda^m_s - \lambda^m_0\big), \qquad \quad \text{ for all  } t < s \leq T,\\
\lambda_t^m&=&\lambda^m.
\end{array}
\label{ODE:1}
\right.
\end{equation}
For $m=1,\ldots,D$, basic theory on ODE provides existence and uniqueness of a solution to \eqref{ODE:1}, which is given by:
\begin{equation}
\lambda^m_s \;\;=\;\; \lambda^m_0 + \big(\lambda^m- \lambda^m_0\big) e^{-D\beta(s-t)}, \quad \text{ for } s \in [t,T], \text{ and } m=1,\ldots, D.
\end{equation}
Since $ \frac{\partial g}{\partial s} = \frac{\partial f}{\partial s} + \sum_{m=1}^{D} \frac{d \lambda^m_s}{ds} \frac{\partial f}{\partial \lambda^m}$, then
$g(t,z)= G(z)-\int_{t}^{T} F(s,z,\lambda_s) \diff s $, which finally leads to the following expression of $\mathcal{T}_1(F)$:
\begin{equation}
\mathcal{T}_1(F)=f(t,z,\lambda) = G(z) - \int_{t}^{T} F \big(s,z,\lambda_s  \big) \diff s .
\label{carT1}
\end{equation}
\noindent Replacing $F$ by $\mathcal{T}_2(f)$ in \eqref{carT1}, we get that $f$ is fixed point of $\mathcal{T}_1 \circ \mathcal{T}_2$ if and only if: 
\[
f(t,\lambda,z) + \sum_{m=1}^{D}\int_{t}^{T} \lambda^m_s f\big(s,z,\lambda_s\big) \diff s = G(z) - \sum_{m=1}^{D} \int_{t}^{T} \lambda^m_s \ssup{a \in A_z} f\big(s,e_m^a(z), \lambda_s+ \alpha_m\big) \diff s .
\]

\noindent Notice $$\frac{\partial f(s, \lambda_s, z) e^{-\sum_{j=1}^{D}\int_{t}^{s} \lambda^j_udu}}{\partial s} =- \sum_{m=1}^{D}\lambda^m_s  e^{-\sum_{j=1}^{D}\int_{t}^{s} \lambda^j_udu} \ssup{a  \in A_z} f\big(s,e_m^a(z), \lambda_s+ \alpha_m\big),$$ so that: 
\begin{align}
f(t,\lambda,z ) &= G(z) e^{-\sum_{m=1}^{D}\int_{t}^{T} \lambda^m_s \diff s } + \sum_{m=1}^{D} \int_{t}^{T} \lambda^m_se^{-\int_{t}^{s}\lambda_udu } \ssup{a \in A_z} f\big(s,e_m^a(z), \lambda_s+ \alpha_m\big) \diff s \nonumber  \\
&=  G(z) e^{-\sum_{m=1}^{D}\int_{t}^{T} \lambda^m_s \diff s }  + \ssup{a \in A_z} \mathbb{E}^a_{t,\lambda,z}\left[ f\big(T_1, Z_1,\lambda_{T_1}+ \alpha_m\big)\right], \label{ope}
\end{align}
where $T_1$ is the first jump time of $N$ larger than $t$, we denote $Z_1=Z_{T_1}$.
Equation \eqref{ope} shows that the fixed point of $\mathcal{T}_1 \circ \mathcal{T}_2$ is characterized as the fixed point of the operator $\mathcal{T}$ defined for any smooth enough function $f$ by: 
\begin{equation}
\label{eq::rewardop}
\mathcal{T}(f)= G(z) e^{-\sum_{m=1}^{D}\int_{t}^{T} \lambda^m_s \diff s }  + \ssup{a  \in A_z} \mathbb{E}^a_{t,\lambda,z}\Big[ f\big(T_1, Z_1,\lambda_{T_1}+ \alpha_m\big)\Big],
\end{equation}
where $\mathbb{E}_{t,\lambda,z}^a[.]$ stands for the expectation conditioned by the events $\lambda_t=\lambda$ and $Z_t=z$, when decision $a$ is taken at time $t$. We recognize here the maximal reward operator of the value function defined in \eqref{eq::valuefunctionPDMDP}. Basic theory on PDMDP shows that the maximal reward operator $\mathcal{T}$ admits $V$ as unique fixed point, which completes step 2.
\qed
\red{\section*{Conclusion}
In this paper, we solved theoretically and numerically a general market-making problem with different microstructural models of order books, by rewriting the problem as a Markov decision process with infinite horizon. This new representation offers a nice and simple characterization of the optimal strategy of market-making, which is implementable relying for example on some quantization and control randomization ideas as proposed in this paper. Others algorithms, like those based on reinforcement learning (see e.g. Chapter 6.5 of \citet{Sutton1998} for an introduction to (deep) $Q$-learning, and/or  \citet{gueant2019deep} for its application to market-making), look particularly well-adapted to solve the market-making problem using its MDP representation, especially in the context of high-dimension. \newline
The proposed methodology can be adapted to solve theoretically and numerically control problems associated to a general class of controlled point processes. }
\red{\section*{Acknowledgment}
We would like to thank the anonymous referees for their valuable comments on the first version of the paper.}

\bibliographystyle{rQUF}
\bibliography{bibtex.bib}

\appendix
\section{Proof of Theorem \ref{theo::Qknn} and Corollary \ref{cor::Qknn}} \label{secappenquantif}

\noindent We divided the proofs of Theorem \ref{theo::Qknn} and Corollary \ref{cor::Qknn} into the following Lemmas.

Lemma \ref{lem::boundepsproj} aims at bounding  the projection error. It relies on \citet{Gyo02}, see p.93, as well as Zador's theorem, stated in Section \ref{sec:zador} for the sake of completeness.

\vspace{2mm}

\begin{lemma}
	\label{lem::boundepsproj} 
	Assume $d \geq 3$, and take $K\;=\;M^{d+2}$ points for the optimal quantization of $\eps_n$, then it holds under \textbf{(H$\mu$)} and \textbf{(HF)}, as $M \to +\infty$,
	\begin{equation}
	\label{ineq:boundProj}
	\eps^{proj}_n \quad = \quad  \mathcal{O}\left(\frac{1}{M^{1/d}} \right),
	\end{equation} where we remind that
	$\eps_{n}^{proj}:= \ssup{a \in A}\lVert  \mathrm{Proj}_{n+1}\left(F(X_n,a,\hat{\varepsilon}_{n})\right)  - F(X_n,a,\hat{\varepsilon}_{n})  \rVert_2$ stands for the average projection error.
\end{lemma}

\begin{proof}   
	Let us take $\eta>0$, and observe that
	\begin{align}
	\mathbb{P}\left(\left| \mathrm{Proj}_{n+1}\big[ F(X_n,a,\hat{\varepsilon}_{n+1})\big]  -F(X_n,a,\hat{\varepsilon}_{n+1}) \right|^2 > \eta  \right) &= \mathbb{E}\left[ \prod_{m=1}^M\mathbb{E}\left[ \1_{\left\vert X_{n+1}^{t,(m)}  - F(X_n,a,\hat{\varepsilon}_{n+1}) \right\vert   > \sqrt{\eta}} \Bigg\vert X_n,\hat{\varepsilon}_{n+1}  \right] \right]\\
	&=\mathbb{E} \left[ \Big(   1- \mu \big[  B\big(  F(X_n,a,\hat{\varepsilon}_{n+1}), \sqrt{\eta}\big)\big]\Big)^M  \right],
	\end{align}
	where for all $x\in E$ and $\eta>0$, $B(x,\eta)$ denote the ball of center $x$ and radius $\eta$.
	Since $x \mapsto (1-x)^M$ is $M$-Lipschitz, we get by application of  Zador's theorem:
	\begin{align}
	\mathbb{P}\left(\left|  \mathrm{Proj}_{n+1}\big[ F(X_n,a,\hat{\varepsilon}_{n+1})\big]   -F(X_n,a,\hat{\varepsilon}_{n+1})\right|^2 > \eta  \right) & \\
	& \hspace{-5cm}\leq M [F]_L [\mu]_L \norme{\hat{\varepsilon}_{n+1}- \varepsilon_{n+1}}_2 + \mathbb{E} \bigg[ \Big(   1- \mu \big(  B\big(  F(X_n,a,\varepsilon_{n+1}), \sqrt{\eta}\big)\big)\Big)^M  \bigg] \\
	& \hspace{-5cm}=  \frac{M[F]_L [\mu]_L}{K^{1/d}} +  \mathbb{E} \left[ \Big(   1- \mu \big(  B\big(  F(X_n,a,\varepsilon_{n+1}), \sqrt{\eta}\big)\big)\Big)^M  \right]  + \mathcal{O} \left( \frac{M}{K^{1/d}} \right),%
	\end{align}
	as   the number of points for the quantization of the exogenous noise $K$ goes to $+\infty$, and where $M$ stands for the size of the grids $\Gamma_{n}$.
	
	Let us introduce $A_1,...,A_{N(\eta)}$, a cubic partition of $\mathrm{Supp}(\mu)$, which is bounded under \textbf{(H$\mu$)}, such that for all $j=1,\ldots,N(\eta)$, $A_j$ has diameter $\eta$. Also, Notice that there exists $c>0$, which only depends on $\mathrm{Supp}(\mu)$, such as
	\begin{equation}
	\label{majoreN}
	N(\eta) \leq \frac{c}{\eta^d}.
	\end{equation}
	If $x \in A_j$, then $A_j \subset B(x,\eta)$, therefore:
	\begin{align}
	\mathbb{E}\Big[ (1-\mu \left( B(X_n,\eta) ) \right)^M  \Big] &= \sum_{j=1}^{N(\eta)} \int_{A_j} \Big(  1-\mu(B(x,\eta)) \Big)^M \mu(dx)\\
	& \leq  \sum_{j=1}^{N(\eta)} \int_{A_j} \Big(  1-\mu(A_j) \Big)^M \mu(dx).   	\label{ineq:lemma1_1}
	\end{align}
	Also notice that:
	\begin{align}
	\label{ineq:lemma1_2}
	\sum_{j=1}^{N(\eta)} \mu(A_j) \Big(  1-\mu(A_j) \Big)^M\leq \sum_{j=1}^{N(\eta)}  \max_z z(1-z)^M \leq \frac{e^{-1} N(\eta)}{M}.
	\end{align}
	Combining \eqref{ineq:lemma1_1} and \eqref{ineq:lemma1_2} leads to 
	\begin{equation}
	\label{ineq:lemma1_3}
	\mathbb{E}\Big[ (1-\mu \left( B(X_n,\eta) ) \right)^M  \Big] \leq \frac{e^{-1} N(\eta)}{M}.
	\end{equation}
	Let $L= 2\lVert \mu \rVert_\infty$ stands for the diameter of the support of $\mu$. We then get, as $M \to +\infty$,
	\begin{align}
	\mathbb{E}\bigg[ \left\vert \mathrm{Proj}_{n+1}\big[ F(X_n,a,\hat{\varepsilon}_{n+1})\big]  -  F(X_n,a,\hat{\varepsilon}_{n+1}) \right\vert^2  \bigg]  &\\
	&\hspace{-5cm}= \int_{0}^{\infty} \mathbb{P}\Big(\left| \mathrm{Proj}_{n+1}\big[ F(X_n,a,\hat{\varepsilon}_{n+1})\big]    -F(X_n,a,\hat{\varepsilon}_{n+1}) \right|^2 > \eta  \Big)\diff \eta \\
	& \hspace{-5cm} \leq \int_{0}^{L^2}  \frac{M[F]_L [\mu]_L}{K^{2/d}}+\mathbb{P}\Big(\abs{ \mathrm{Proj}_{n+1}\big[ F(X_n,a,\hat{\varepsilon}_{n+1})\big]   -F(X_n,a,\epsilon_{n+1})} > \sqrt{\eta}  \Big)\diff \eta \\
	& \hspace{-5cm}=  \int_0^{L^2} \min \left( 1, \frac{e^{-1} N(\sqrt{\eta})}{M} \right) \diff \eta +\mathcal{O}\left(\frac{M}{K^{1/d}} \right) \\
	& \hspace{-5cm}= \int_0^{L^2} \min \left( 1, \frac{ c \eta^{-d/2}}{eM} \right) \diff \eta  +\mathcal{O}\left(\frac{M}{K^{1/d}} \right) \\
	& \hspace{-5cm}= \int_0^{(c/(eM))^{(2/d)}} 1 d\eta + \int_{(c/(eM))^{(2/d)}}^{L^2} \frac{ c \eta^{-d/2}}{eM}  \diff \eta +\mathcal{O}\left(\frac{M}{K^{1/d}} \right) \\
	& \hspace{-5cm}= \frac{\tilde{c}^2}{M^{2/d}} +\mathcal{O}\left(\frac{M}{K^{1/d}} \right), \label{ineq:proj1}
	\end{align} 
	where $\tilde{c}$ is defined as $\tilde{c}:= \sqrt{\frac{d}{d-2}} \left( \frac{c}{e} \right)^{1/d}$, and where we used \eqref{ineq:lemma1_3} and \eqref{majoreN} to go from the second to the third line. 
	It remains to take $K=M^{d+1}$ points for the optimal quantization of the exogenous noise, and then take square root of equality \eqref{ineq:proj1}, in order to derive \eqref{ineq:boundProj}. 
\end{proof}

\vspace{3mm}

\begin{lemma}
	\label{lem:majoreProjErrorX}
	Assume $d \geq 3$, take $K\;=\;M^{d+2}$ points for the optimal quantization of $\eps_n$,	and let $x \in E$. Then it holds under \textbf{(H$\mu$)} and \textbf{(HF)}, as $M \to +\infty$:
	\begin{equation}
	\label{ineq:boundProj_2}
	\eps^{proj}_n(x) \quad = \quad  \mathcal{O}\left(\frac{1}{M^{1/d}} \right),
	\end{equation} where $\eps_{n}^{proj}(x)$, defined as
	$\eps_{n}^{proj}(x):= \ssup{a \in A}\lVert  \mathrm{Proj}_{n+1}\left(F(x,a,\hat{\varepsilon}_{n})\right)  - F(x,a,\hat{\varepsilon}_{n})  \rVert_2$, stands for the later-projection error at state $x$.\\
\end{lemma}
\begin{proof}
	Following the same steps as those used to prove Lemma \ref{lem::boundepsproj}, we show that:
	\begin{align}
	\mathbb{P}\left(\left|  \mathrm{Proj}_{n+1}\big[ F(x,a,\hat{\varepsilon}_{n+1})\big]   -F(x,a,\hat{\varepsilon}_{n+1})\right|^2 > \eta  \right) & \\
	& \hspace{-5cm}=  \frac{M[F]_L [\mu]_L}{K^{1/d}} +  \mathbb{E} \left[ \Big(   1- \mu \big(  B\big(  F(x,a,\varepsilon_{n+1}), \sqrt{\eta}\big)\big)\Big)^M  \right]  + \mathcal{O} \left( \frac{M}{K^{1/d}} \right),
	\end{align}
	as $K \to +\infty$, and moreover, $$\mathbb{E} \left[ \Big(   1- \mu \big(  B\big(  F(x,a,\varepsilon_{n+1}), \sqrt{\eta}\big)\big)\Big)^M  \right] \leq \frac{e^{-1} N(\eta)}{M},$$ holds,  which is enough to complete the proof of Lemma \ref{lem:majoreProjErrorX}.
\end{proof}

\vspace{3mm}
\begin{lemma}
	\label{lem:lip}
	Under \textbf{(HF)}, for $n=0,\ldots,N$ there exists constant $\left[\hat{V}_n^Q\right]_L>0$ such that for $x, x' \in E$, it holds as $M \to \infty$:
	\begin{equation}
	\label{eq:VQalmostLip}
	\left|  \hat{V}_n^Q(x) -  \hat{V}_n^Q(x')  \right|\; \leq  \;  \left[\hat{V}_n^Q\right]_L \left|   x-x'\right| + \mathcal{O} \left( \frac{1}{M^{1/d}}\right).
	\end{equation}
	Moreover, following bounds holds on $\left[\hat{V}_n^Q\right]_L$, for $n=0,\ldots,N$:
	\begin{equation}
	\begin{cases}
	\left[\hat{V}_N^Q\right]_L &\leq [g]_L \\
	\left[\hat{V}_n^Q\right]_L &\leq [f]_L + [F]_L \left[\hat{V}_{n+1}^Q\right]_L, \qquad \text{ for } n=0,...,N-1.
	\end{cases}
	\label{ineq:constlipVQ}
	\end{equation}
\end{lemma}
\begin{proof}  Let us show that by induction that $\hat{V}_N^Q$ is Lipschitz. 
	First, notice that \eqref{eq:VQalmostLip} holds at terminal time $n=N$, if one define $\left[\hat{V}_N^Q\right]_L$ as $ \left[\hat{V}_N^Q\right]_L = [g]_L$ . 
	Let us take   $x,x' \in E$.  	Assume $\left|  \hat{V}_{n+1}^Q(x) -  \hat{V}_{n+1}^Q(x')  \right|\; \leq  \;  \left[\hat{V}_{n+1}^Q\right]_L \left|   x-x'\right| + \mathcal{O} \left( \frac{1}{M^{1/d}}\right)$ holds for some $n=0,\ldots,N-1$. Let us show that 	
	\begin{equation}
	\left|  \hat{V}_n^Q(x) -  \hat{V}_n^Q(x')  \right|\; \leq \;  \left[\hat{V}_n^Q\right]_L \left|   x-x'\right| + \mathcal{O} \left( \frac{1}{M^{1/d}}\right),
	\end{equation}
	where $\left[\hat{V}_n^Q\right]_L$ is defined in \eqref{ineq:constlipVQ}. Notice that, by the dynamic programming principle and the triangular inequality, it holds:
	\begin{align}
	\vert \hat{V}_n^Q(x)-\hat{V}^Q_n(x') \vert & \leq [f]_L\left| x-x' \right| \\
	& \hspace{-1cm}+ \ssup{a} \mathbb{E}^a_n\left[ \left\vert  \hat{V}_{n+1}^Q\big(\mathrm{Proj}_{n+1} \left(F(x,a,\hat{\eps}_{n+1})\right)\big) - \hat{V}_{n+1}^Q\big(\mathrm{Proj}_{n+1} \left(F(x',a,\hat{\eps}_{n+1})\right)\big) \right\vert \right]   \\
	&  \hspace{-1.4cm} \leq  [f]_L\left| x-x' \right| + \left[\hat{V}_{n+1}^Q\right]_L \ssup{a}  \mathbb{E} \left[ \left\vert \mathrm{Proj}_{n+1} \left(F(x,a,\hat{\eps}_{n+1})\right) - F(x,a,\hat{\eps}_{n+1}) \right\vert \right] \\
	&  \hspace{-1cm}   + \mathcal{O} \left(\frac{1}{M^{1/d}}\right) \\
	&  \hspace{-1.4cm} \leq  \left( [f]_L + \left[\hat{V}_{n+1}^Q\right]_L [F]_L \right) \left| x-x' \right| + \mathcal{O} \left(\frac{1}{M^{1/d}}\right) \\
	&  \hspace{-1.4cm} \leq \left[\hat{V}_{n}^Q\right]_L \left| x-x' \right| + \mathcal{O} \left(\frac{1}{M^{1/d}}\right),
	\end{align}
	which completes the proof of \eqref{eq:VQalmostLip}.
\end{proof}

\vspace{3mm}
\noindent We now proceed to the proof of Theorem \ref{theo::Qknn}.
\begin{proof} (of Theorem \ref{theo::Qknn})
	Combining inequality $
	\left \vert u_1+ u_2+ u_3 \right\vert^2 \leq 3  \left( \vert u_1 \vert^2 +  \vert u_2 \vert^2 + \vert u_3 \vert^2 \right)
	$ that holds for all $u_1,u_2,u_3 \in \R$ with inequality $ \left\vert \ssup{i \in I} a_i - \ssup{i \in I } b_i \right\vert \leq \ssup{i \in I} \vert a_i - b_i \vert$  that holds for all families $(a_i)_{i \in I}$ and $(b_i)_{i \in I}$ of reals, and all set $I$, we have:
	\begin{align}
	\lVert \hat{V}^Q_n(X_n)- V_n(X_n) \rVert_2^2  &\leq 3 \; \mathbb{E}\Bigg[ \ssup{a \in A} \mathbb{E}_{n,X_n} \left\vert \hat{V}^Q_{n+1}\left(\mathrm{Proj}_{n+1}\left( F(X_n,a,\hat{\varepsilon}_{n+1})\right) \right)-  \hat{V}^Q_{n+1}(F(X_n,a,\hat{\varepsilon}_{n+1})) \right\vert^2 \nonumber \\
	& \hspace{1cm}+ \ssup{a \in A} \mathbb{E}_{n,X_n} \left\vert \hat{V}^Q_{n+1}(F(X_n,a,\hat{\varepsilon}_{n+1})) - \hat{V}^Q_{n+1}(F(X_n,a,\varepsilon_{n+1})) \right\vert^2 \\
	&\hspace{1cm} + \ssup{a \in A} \mathbb{E}_{n,X_n} \left\vert \hat{V}^Q_{n+1}(F(X_n,a,\varepsilon_{n+1})) - V_{n+1}(F(X_n,a,\varepsilon_{n+1})) \right\vert^2  \Bigg]
	\end{align}
	where $\mathbb{E}_{n,X_n} $ stands for the expectation conditioned by the state $X_n$ at time $n$.  It holds as $M\to +\infty$, using Lemma \ref{lem:lip}:
	\begin{align}
	\lVert \hat{V}^Q_n(X_n)- V_n(X_n) \rVert_2^2 &\leq 3 \; \left[\hat{V}_n^Q\right]_L \mathbb{E}\Bigg[ \ssup{a} \mathbb{E}_{n,X_n} \left[\vert \mathrm{Proj}_{n+1}\left(F(X_n,a,\hat{\varepsilon}_{n+1})\right)-  F(X_n,a,\hat{\varepsilon}_{n+1}) \vert^2 \right] \nonumber\\
	& \hspace{2.5cm}+ \ssup{a} \mathbb{E}_{n,X_n} \left[ \vert F(X_n,a,\hat{\varepsilon}_{n+1}) - F(X_n,a,\varepsilon_{n+1}) \vert^2   \right] \Bigg] \nonumber\\
	&\hspace{0.4cm} + 3 \;\; \lVert r \rVert_\infty  \mathbb{E}\Big[  \vert \hat{V}^Q_{n+1}(X_{n+1})) - V_{n+1}(X_{n+1})) \vert^2  \Big] +\mathcal{O}\left(\frac{1}{M^{1/d}}\right)  \label{ineq::Qknn}
	\end{align}	
	Under \textbf{(HF)}, \eqref{ineq::Qknn} can then be rewritten as:
	\begin{align}
	\lVert \hat{V}^Q_n(X_n)- V_n(X_n) \rVert_2^2 \;\; &\leq  \;\; 3 \left[\hat{V}_n^Q\right]_L \Big( [F]_L^2 (\epsilon_{n}^Q)^2 + (\epsilon^{proj}_n)^2  \Big)  \\
	&\hspace{.4cm} \;\;+ 3 \lVert r \rVert_\infty \lVert \hat{V}^Q_{n+1}(X_{n+1})- V_{n+1}(X_{n+1}) \rVert_2^2 + \mathcal{O}\left(\frac{1}{M^{1/d}}\right).
	\end{align}
	\eqref{ineq:theo2} then follows by induction, which completes the proof of Theorem \ref{theo::Qknn}.
\end{proof}

\vspace{3mm}
\begin{proof} (of Corollary \ref{cor::Qknn})\\
	Corollary \ref{cor::Qknn} is straightforward by plugging the bound for the projection error provided by Lemma \ref{lem::boundepsproj} and the one of the quantization error provided by the Zador's Theorem into \eqref{ineq:theo2}.
\end{proof}

\section{Zador's Theorem}
\label{sec:zador}

\begin{theorem}[Zador's theorem]
	\label{thm:Zador}
	Let us take $n=0,\ldots,N$,  and denote by $K$ the number of points for the quantization of the exogenous noise $\varepsilon_n$.\\
	Assume that $\mathbb{E} \left[ \vert \varepsilon_n\vert ^{2+\eta} \right] < +\infty$ for some $\eta>0$. Then, there exists a universal constant $C>0$ such that:
	\begin{equation}
	\lim_{M \to +\infty} \Big( M^{\frac{1}{d}} \lVert \hat{\varepsilon}_n - \varepsilon_n \rVert_2\Big) =C
	\end{equation}
\end{theorem}

\begin{proof}
	We refer to \citet{GrLu00} for a proof of Theorem \ref{thm:Zador}.
\end{proof}

\end{document}